\documentclass{aa}
\usepackage{graphicx}               
\usepackage{placeins}               
\graphicspath{{figures/}}           
\usepackage{txfonts}                
\usepackage{lscape}                 
\usepackage[
	colorlinks=true,
	urlcolor=cyan,
	linkcolor=blue,
	citecolor=blue
]{hyperref}
\usepackage{comment}
\usepackage{multirow}
\usepackage{natbib}
\bibpunct{(}{)}{;}{a}{}{,}          
\defcitealias{jwsttransitingexoplanetcommunityearlyreleasescienceteam_2023}{JTERS23}

\usepackage{amsmath}
\usepackage{mathabx}
\usepackage{siunitx}
\usepackage[version=4]{mhchem}
\DeclareSIUnit{\solarradius}{R_\sun}
\DeclareSIUnit{\solarmass}{M_\sun}
\DeclareSIUnit{\jupiterradius}{R_J}
\DeclareSIUnit{\jupitermass}{M_J}
\DeclareSIUnit{\au}{au}
\DeclareSIUnit{\gauss}{G}
\DeclareSIUnit{\erg}{erg}
\DeclareSIUnit{\year}{yr}
\DeclareSIUnit{\day}{d}
\DeclareSIUnit{\dex}{dex}
\DeclareSIUnit{\bar}{bar}

\usepackage[normalem]{ulem}
\usepackage{color}

\begin{document}

    \title{Knobs and dials of retrieving JWST transmission spectra}
	\subtitle{I. The importance of p-T profile complexity}
	
	\author{
		S. Schleich \inst{1}
		\and
		S. Boro Saikia \inst{1}
		\and
		Q. Changeat \inst{2,3,4}
		\and
		M. Güdel \inst{1}
		\and
		A. Voigt \inst{5}
		\and
		I. Waldmann \inst{3}
	}
	
	\institute{
		Department of Astrophysics, 
		University of Vienna,
		Türkenschanzstrasse 17, 1180 Vienna, Austria\\
		\email{simon.schleich@univie.ac.at}
		\and		
        Kapteyn Astronomical Institute, 
        University of Groningen, PO Box 800, 
        9700 AV Groningen, The Netherlands
        \and
		European Space Agency (ESA), ESA Office, 
		Space Telescope Science Institute (STScI), 
		Baltimore MD 21218, USA
		\and
		Department of Physics and Astronomy, 
		University College London, 
		Gower Street, WC1E 6BT London, United Kingdom
		\and
		Department of Meteorology and Geophysics, 
		University of Vienna,
		Josef-Holaubek-Platz 2, Vienna, Austria
	}
	
	\date{Received XXX; Accepted YYY}
	
	
	\abstract
    {
        When retrieving exoplanet atmospheric characteristics from spectroscopic observations, parameter estimation results are strongly depend on the chosen forward model. In the era of the \textit{James Webb} Space Telescope (JWST) and other next-generation facilities, the increased signal-to-noise (S/N), wavelength coverage, and spectral resolution of observations warrant closer investigations into factors that could inadvertently bias the results of these retrievals.
	}
	{
        We investigate the impact of utilising multipoint pressure-temperature (p-T) profiles of varying complexity on the retrieval of synthetically generated hot Jupiter transmission spectra modelled after state-of-the-art observations of the hot Jupiter WASP-39~b with JWST.
	}
	{
        We perform homogenised atmospheric retrievals with the \texttt{TauREx} retrieval framework on a sample of synthetically generated transmission spectra, accounting for varying cases of underlying p-T profiles, cloud-top pressures, and expected noise levels. These retrievals are performed using a fixed-pressure multipoint p-T prescription with increasing complexity, ranging from isothermal to an eleven-point profile. We evaluate the performance of the retrievals based on the Bayesian model evidence, and the accuracy of the retrievals compared to the known input parameters.
	}
	{
	    We find that performing atmospheric retrievals using an isothermal prescription for the pressure-temperature profile consistently results in wrongly retrieved atmospheric parameters when compared to the known input parameters. For an underlying p-T profile with a fully positive lapse rate, we find that a two-point profile is sufficient to retrieve the known atmospheric parameters, while under the presence of an atmospheric temperature inversion, we find that a more complex profile is necessary.
	}
	{
        Our investigation shows that, for a data quality scenario mirroring state-of-the-art observations of a hot Jupiter with JWST, an isothermal p-T prescription is insufficient to correctly retrieve the known atmospheric parameters. We find a model complexity preference dependent on the underlying pressure-temperature structure, but argue that a p-T prescription on the complexity level of a four-point profile should be preferred. This represents the overlap between the lowest number of free parameters and highest model preference in the cases investigated in this work.
    }
	%
	
	%
	%
	

	\maketitle
	
	\section{Introduction}\label{sec:introduction}
		
		In the three decades since their first discovery \citep{wolszczan_1992,mayor_1995}, the inventory of known extrasolar planets has been significantly expanded, and now counts more than \num{5700} confirmed objects\footnote{NASA Exoplanet Archive (\url{https://exoplanetarchive.ipac.caltech.edu/}), 29 July 2024}. These vary greatly in their characteristics from small, dense worlds to inflated giant planets, and encompass a parameter regime beyond the planets found in the Solar System. Techniques used to detect exoplanets vary, but among them the most prolific method to date is the ``transit method'', where the amount of incident stellar flux received by an observer is reduced by the transit of an exoplanetary companion obscuring part of the stellar disk, which allows the determination of an effective size of the planet.
		
		Studying these transit events provides a unique window into the nature of exoplanetary atmospheres. During the primary transit, the observed stellar flux passes through the atmospheric terminator, where it is attenuated by opacity sources such as molecules or condensates, influencing the observed size of the planet. The wavelength-dependent behaviour of this characteristic ``depth'' of the transit, which is proportional to the square of the apparent planetary radius in units of the host stars radius, therefore depends on the nature of the underlying exoplanetary atmosphere. This fact was used in the first successful detections of absorption caused by atomic \citep{charbonneau_2002,vidal-madjar_2003} and molecular \citep{barman_2007,tinetti_2007} species in the atmospheres of inflated gas giants, called ``hot Jupiters''. Since these initial results, the inventory of characterised exoplanetary atmospheres has increased significantly. The preeminent space-based observatories that have facilitated the characterisation of exoplanet atmospheres are the \textit{Hubble} Space Telescope (HST) and the \textit{Spitzer} Space Telescope, which have predominantly provided access to the atmospheres of hot Jupiters. We refer to, for instance, \citet{kreidberg_2018} and \citet{madhusudhan_2019} for comprehensive reviews on the inventory of characterised exoplanet atmospheres.
  
        The launch of the James Webb Space Telescope (JWST, \citealt{gardner_2006,gardner_2023}) has already started to provide a significant leap forward in our ability to characterise exoplanet atmospheres. One of the first investigations of the atmosphere of a hot Jupiter using the spectroscopic capabilities of JWST has revealed spectral signatures associated with the presence of \ce{CO2} (\citealt{jwsttransitingexoplanetcommunityearlyreleasescienceteam_2023}, hereby referred to as \citetalias{jwsttransitingexoplanetcommunityearlyreleasescienceteam_2023}). Subsequent observational campaigns have lead to the successful atmospheric characterisation of additional hot Jupiters \citep[e.g.][]{august_2023,bean_2023,bell_2023,dyrek_2024}. JWST has also improved the capability of investigating the atmospheres of smaller planets, such as sub-Neptunes \citep[e.g.][]{kempton_2023,madhusudhan_2023,benneke_2024}, and detecting atmospheres around terrestrial planets \citep[e.g.][]{lustig-yaeger_2023,moran_2023,zieba_2023,kirk_2024}. Together with future missions such as Ariel \citep{tinetti_2018}, this leap in accessible exoplanet atmospheres will permit large-scale comparisons of system-level, as well as population-level parameters.
  
        To determine exoplanet atmospheric characteristics, such as the pressure-temperature (p-T) structure and chemical composition based on spectroscopic observations, an inversion process referred to as ``atmospheric retrieval'' has found broad application \citep{madhusudhan_2009}. With this method, the preferred parameter values of an atmospheric forward model are determined through a data-guided statistical sampling process. A variety of atmospheric retrieval codes have been developed that use different parameter exploration methods, including NEMESIS \citep{irwin_2008,lee_2012}, petitRADTRANS \citep[pRT,][]{molliere_2019,nasedkin_2024}, and \texttt{TauREx} \citep{waldmann_2015-1,al-refaie_2021}. We refer to \citet{macdonald_2023} for a comprehensive overview of atmospheric retrieval framework.

        While the statistical sampling of model parameters in atmospheric retrieval is guided by the observational data through Bayesian inference, the reported results of atmospheric characterisation processes also significantly depend on the model parameters themselves. This is exemplified by the varying approaches taken to account for the molecular constituent profiles of atmospheric species in forward models. ``Free'' chemistry is a heuristic approach, fitting atmospheric volume-mixing ratios (VMRs) individually for each considered species, coarsely determining the presence and abundance values of atmospheric absorbers through a large number of free parameters \citep[e.g.][]{changeat_2019,al-refaie_2022}. Self-consistent chemical models rely on an underlying chemical network, and are commonly derived under the assumption of thermochemical equilibrium to infer atmospheric molecular profiles through the minimisation of Gibbs-free energy \citep[e.g.][]{stock_2018,woitke_2018,agundez_2020}. While these models depend on few free parameters, they are also restricted to their model assumptions.

        The p-T profile is another important characteristic in the context of exoplanet atmospheres. Measurements within the Solar System have illustrated the complex pressure-temperature structure of atmospheres accessible to in situ investigations \citep[e.g.][]{seiff_1998,fulchignoni_2005,koskinen_2015,limaye_2017}, and underline the fact that characterisation results for exoplanets strongly depend on the assumptions of simplified models. In atmospheric retrieval, p-T profiles represent another aspect of forward models with a range of underlying assumptions. Analogous to the chemical profiles, a ``free'' p-T profile retrieves temperature (and potentially associated pressure) values within the atmospheric domain on a purely heuristic basis. The simplest form of this is an isothermal (or one-point) prescription, which has a pressure-independent temperature value as its only free parameter. Multipoint p-T profiles allow for a large degree of flexibility in the determination of the atmospheric temperature structure at the cost of an increasingly large number of free parameters, especially if both the pressure node location as well as the associated temperature are chosen as free parameters. The other commonly used method of implementing the p-T structure into atmospheric forward models are parametric p-T profiles \citep[e.g.][]{madhusudhan_2009,guillot_2010}. Retrieving information about the p-T structure of an exoplanetary atmosphere is more commonly associated with the analysis of emission spectra, which probe large pressure regimes of the day-side of exoplanetary atmospheres. \citet{line_2012,line_2013} have investigated the impact of using different parametric and heuristic p-T prescriptions on the retrieval of synthetic emission spectra of hot Jupiters, showing a signal-to-noise (S/N) and spectral resolution dependent preference for the use of a parametric profile. \citet{blecic_2017} compared the performance of different parametric p-T profiles in retrievals of synthetic hot Jupiter emission spectra generated from 3D radiative-hydrodynamic simulations.

        Compared to this, the retrieval of atmospheric spectra observed during a primary transit are more commonly associated with being sensitive to the characterisation of molecular constituents, and the signatures of clouds and hazes. An isothermal prescription for the atmospheric terminator region probed in primary transit observations has found ample application in the analysis of HST data \citep[e.g.][]{tsiaras_2018,roudier_2021}. However, \citet{rocchetto_2016} have shown that for a synthetic, cloud-free hot Jupiter transmission spectrum, under the assumption of JWST-equivalent quality with larger wavelength coverage and increased spectra resolution, the implementation of an over-simplified p-T prescription can significantly bias the retrieved atmospheric VMRs, when compared to a retrieval using a parametric p-T profile. A comprehensive review on outstanding challenges in atmospheric retrievals can be found in, for instance, \citet{barstow_2020}.
        
        In this work, we systematically investigate the influence of p-T prescriptions of increasing complexity on atmospheric retrievals of synthetically generated transmission spectra. We quantify the drawbacks of using an over-simplified, isothermal assumption in connection with the transmission spectrum of a hot Jupiter of the quality produced by JWST. We model these spectra after the recent observations of the planet WASP-39~b, accounting different assumptions about the atmospheric p-T profile, cloud coverage, and transit-depth uncertainty. Furthermore, we determine a feasible level of complexity for the choice of a heuristic p-T profile in the retrieval process, investigating different assumptions about the underlying atmospheric characteristics.

    
    \section{Methods}\label{sec:methods}

        To generate atmospheric forward models and perform atmospheric retrievals, we make use of the open-source atmospheric retrieval framework \verb|TauREx| \citep{waldmann_2015-1, al-refaie_2021}, specifically the version \verb|TauREx3.1|\footnote{\url{https://github.com/ucl-exoplanets/TauREx3_public}} \citep{al-refaie_2022}. \verb|TauREx| performs radiative transfer calculations for transit or eclipse geometries, as well as atmospheric parameter retrieval as a fully-Bayesian parameter inference network. \verb|TauREx| has found extensive application in the retrieval of exoplanet atmospheric properties, including the characterisation of transmission \citep[e.g.][]{tsiaras_2018,edwards_2021,gressier_2022,saba_2022,edwards_2023} and emission \citep[e.g.][]{changeat_2022,edwards_2024} spectra, encompassing a variety of planetary types, ranging from Earth-sized planets to hot Jupiters. To generate exoplanet transmission spectra from atmospheric forward models, we make use of absorption cross-sections from the ExoMol project \citep{tennyson_2020, chubb_2021}, as well as from the HITRAN \citep{gordon_2022} and HITEMP \citep{rothman_2010} archive, and from the Rayleigh scattering incorporated in \verb|TauREx| \citep{cox_2015}. We list individually used opacity contributions in Table~\ref{tab:atm-opacity-data}. To perform parameter estimations, we make use of nested sampling through \verb|MultiNest| \citep{feroz_2009, buchner_2014}, with a standardised value for the evidence tolerance of \num{0.5}, and \num{700} live points \citep{feroz_2009}, in all cases.
        
        To investigate the possibility of recovering more complex pressure-temperature structures from primary transit observations taken by JWST, we perform atmospheric retrievals on a set of synthetic transmission spectra. We model our synthetic dataset after the NIRSpec PRISM observation of the hot Jupiter WASP-39~b (PID: 1366, PI: N. Batalha, Co-PIs: J. Bean and K. Stevenson), to mimic the transmission spectrum of an inflated hot Jupiter. In the construction of atmospheric forward models, we use different assumptions about the cloud deck, underlying p-T profile, and associated precision of the transmission spectrum. Using these forward models as observational data, we perform atmospheric retrievals using a range of multipoint p-T profiles with fixed pressure nodes, ranging between an isothermal (or one-point) prescription, and an eleven-point profile. The latter represents a self-retrieval validation process, as a fixed-pressure eleven-point profile was used to generate the synthetic input spectra (we refer to Sect.~\ref{ssec:synthetic-pt} for more details). In all cases (including the setup of the synthetic transmission spectra), the p-T profile is defined through fixed pressure-temperature nodes. The atmospheric p-T profile is then created using a smoothed linear interpolation on the atmospheric pressure grid, using a standard smoothing window of ten atmospheric layers. We evaluate the preference of different model setups based on the reported Bayesian evidence values, as well as their performance in retrieving the correct underlying atmospheric parameter values.

        \subsection{Synthetic input data}\label{sec:mock-data}

			We generate synthetic atmospheric transmission spectra using the planetary and stellar parameters given in Table~\ref{tab:system-pars} to mimic a hot Jupiter system akin to WASP-39~b. We define the extent of the atmospheric pressure domain used in these forward models as $\log_{10}(p\,[\mathrm{bar}]) \in \left[1; -9\right]$, using \num{110} layers. As a standardised setup for all cases, we construct a primary, \ce{H2 / He}-dominated atmosphere, and additionally include \ce{H2O}, \ce{CO}, \ce{CO2}, \ce{H2S}, and \ce{CH4} as atmospheric absorbers, with constant VMRs throughout the atmosphere. Their associated values are given in Table~\ref{tab:synth-pars}. We also include contributions from collision-induced absorption (CIA) of \ce{H2-H2} and \ce{H2-He} (see Table~\ref{tab:atm-opacity-data} for references of individual opacity data sources), and Rayleigh scattering \citep{cox_2015}. 
   
            We bin the high-resolution forward models generated with \verb|TauREx| to the resolution of NIRSpec PRISM instrument configuration. In this process, we take into consideration that the full exposure of the NIRSpec PRISM observation of WASP-39~b (after 5 groups) is saturated at wavelengths below approximately \SI{2.2}{\micro\meter}. Therefore, we restrict the synthetic transmission spectra generated in this work to the same wavelength regime to mimic the information content present in the actual observational data (we refer to \citetalias{jwsttransitingexoplanetcommunityearlyreleasescienceteam_2023} and \citealt{rustamkulov_2023}, as well as Schleich et al. (in prep., henceforth referred to as Paper II) for more details about the saturated region of this spectrum). The sample of synthetic transmission spectra  used as the input data set is differentiated in three aspects -- the underlying pressure-temperature profile (two cases), the cloud-top pressure of a flat-opacity cloud deck (three cases), and the precision of the transit depth values (two cases). In total, this results in twelve distinct atmospheric forward models.

    		\begin{table}
    			\caption{WASP-39 system parameters}
    			\label{tab:system-pars}
    			\centering
    			\renewcommand{\arraystretch}{1.5}
    			\begin{tabular}{c c c}
    				\hline\hline
    				Parameter & Value & Unit \\
    				\hline
    				\multicolumn{3}{c}{\textbf{Star}} \\
    				\hline
    				$M_\mathrm{*}$ & $0.918 \pm 0.047$ & \si{\solarmass} \\
    				$R_\mathrm{*}$ & $1.013 \pm 0.022$ & \si{\solarradius} \\
    				$T_\mathrm{eff}$ & $5485 \pm 50$ & \si{\kelvin} \\
    				$[\ce{Fe/H}]$ & $0.01 \pm 0.09$ & \si{\dex} \\
    				$\log g$ & $4.41 \pm 0.15$ & \si{\dex} \\
    				\hline
    				\multicolumn{3}{c}{\textbf{Planet}} \\
    				\hline
    				$M_\mathrm{p}$ & $0.281 \pm 0.032$ & \si{\jupitermass} \\
    				$R_\mathrm{p}$ & $1.279 \pm 0.040$ & \si{\jupiterradius} \\
    				$P$ & $4.0552941 \pm \num{3.4d-6}$ & \si{\day} \\
    				\hline
    			\end{tabular}
    			\tablefoot{Stellar and planetary parameters and uncertainties are taken from \citet{mancini_2018}.}
    		\end{table}

			\begin{table}
				\caption{Atmospheric species for synthetic forward models}
				\label{tab:synth-pars}
				\centering
				\renewcommand{\arraystretch}{1.5}
				\begin{tabular}{cc}
					\hline\hline
					Parameter 		& Value \\
					\hline
					$\ce{He / H2}$ 	& 0.13 \\
					$X_{\ce{H2O}}$ 	& \SI{5d-3}{} \\
					$X_{\ce{CO}}$ 	& \SI{4d-3}{} \\
					$X_{\ce{H2S}}$ 	& \SI{5d-4}{} \\
					$X_{\ce{CO2}}$ 	& \SI{5d-5}{} \\
					$X_{\ce{CH4}}$ 	& \SI{1d-7}{} \\
					\hline
				\end{tabular}
				\tablefoot{
					Values for the volume-mixing ratios, $X_\mathrm{M}$, are inspired by the results presented in \citetalias{jwsttransitingexoplanetcommunityearlyreleasescienceteam_2023}. We refer to Table~\protect\ref{tab:atm-opacity-data} for the sources of these opacities.
				}
			\end{table}

			We point out that in the construction of the synthetic transmission spectra, we do not scatter our spectral data points. As we are interested in investigating the potential biases of using simplified pressure-temperature characterisations in the retrieval of atmospheric properties, added randomisation through a self-scattering step of the spectrum could introduce additional, non-controllable biases into the retrieval results. We therefore use the non-scattered instance of the transmission spectrum as an idealised case representing the averaging of results from a larger set of randomised instances \citep[e.g.][]{feng_2018,changeat_2019}.

			\subsubsection{Pressure-temperature profiles}\label{ssec:synthetic-pt}

\begin{figure}
					\centering
					\includegraphics[width=0.6\hsize]{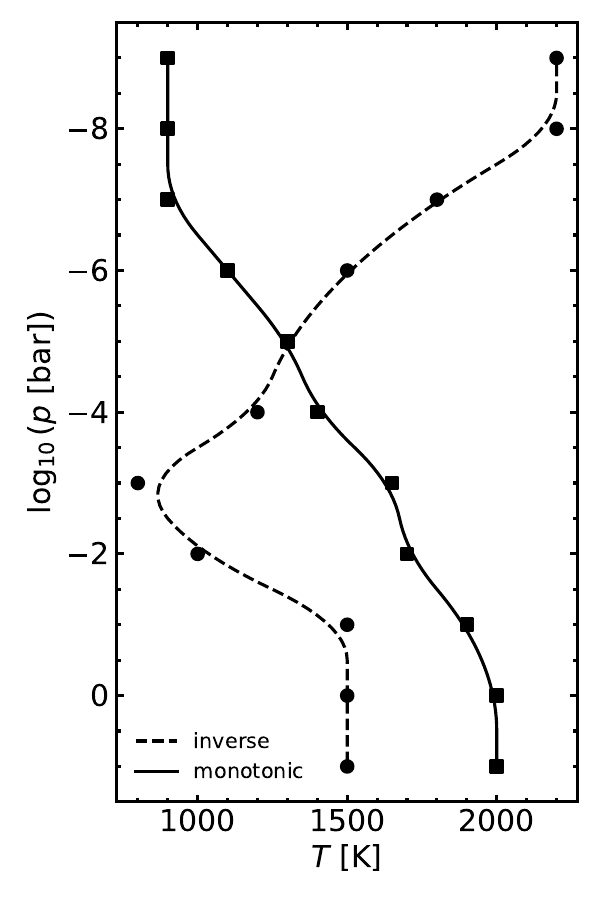}
					\caption{Constructed pressure-temperature profiles, which are used to create synthetic transmission spectra for our test cases. Temperature (in \si{\kelvin}) is shown on the x-axis, and pressure (in \si{\bar}) on the y-axis. The ($p$, $T$)-pairs (filled markers) are log-uniformly distributed in the pressure regime of the atmosphere. The shape of the two profiles characterise a simpler, purely positive temperate lapse rate (solid line), and a more complex structure including a temperature inversion point (dashed line). We note that the ($p$, $T$)-pair at $p = 10^{-5}$ \si{\bar} is the same for both cases.}
					\label{fig:pT-GT}
				\end{figure}

				We investigate two cases of the underlying pressure-temperature structure as reference cases. These are shown in Fig.~\ref{fig:pT-GT} and categorised as a less complex profile, represented by a positive temperature lapse rate (i.e. a negative temperature gradient with increasing altitude, referred to as ``monotonic'' from here on), and a more complex case characterised by a temperature inversion between a positive temperature lapse rate in the lower-, and negative temperature lapse rate in the upper atmosphere (referred to as ``inverse'' from here on). Both cases are constructed as eleven-point p-T profiles, where the pressure nodes are log-uniformly distributed in the pressure domain of our atmospheric model. The shapes of the ``monotonic'' and ``inverse'' profile are inspired by the cases of the hot Jupiters WASP-76~b and WASP-77~A~b from \citet{changeat_2022}, respectively. We use these two profile shapes to investigate the performance of an isothermal prescription, as well as multipoint prescriptions of the p-T profile for two cases of varying complexity. The synthetic transmission spectra generated with the ``monotonic'' and ``inverse'' input pressure-temperature profile are shown in the top and middle panel of Fig.~\ref{fig:synthetic-cases}, respectively.

			\subsubsection{Atmospheric cloud deck}\label{ssec:cloud-cases}

				We construct the synthetic transmission spectra under the assumption of a flat-opacity cloud deck, defined as an optical depth step-function,
				\begin{equation}
					\tau(\lambda, z) = 
					\begin{cases}
						\infty	&\quad \text{if } p(z) \geq p_\mathrm{cloud} \\
						0		&\quad \text{if } p(z) < p_\mathrm{cloud},
					\end{cases}
				\end{equation}
				where $\tau(\lambda, z)$ denotes the optical depth at a certain wavelength, $\lambda$, and altitude, $z$, $p(z)$ the pressure at altitude $z$, and $p_\mathrm{cloud}$ the designated cloud-top pressure. \citet{sing_2016} have shown a distribution from clear to cloudy atmospheres in a comparative analysis of the transmission spectra of ten hot Jupiters, where clouds and hazes mute spectral features in the respective atmospheres. To cover a wide range of possible modulations to the transmission spectrum, we include three different cases of the cloud-top pressure, more specifically $\log_{10}(p_\mathrm{cloud} \, [\mathrm{bar}]) \in \left\{0, -3, -5\right\}$. The influence of these cloud decks on the final transmission spectrum is illustrated in the first two panels of Fig.~\ref{fig:synthetic-cases}. In the case of a low-pressure (or high-altitude) cloud deck, the molecular absorption features show significantly lower amplitudes, which are much more pronounced in the case of a high-pressure (or low-altitude) cloud deck.

			\subsubsection{Transit depth precision}

				The third varying parameter we include in our model spectra is the precision of the transmission spectrum, $\sigma_\mathrm{td}$, which we generate in two different ways. The first one uses error bars generated with the \verb|PandExo| \citep{batalha_2017}, a noise-simulation tool for observations with JWST. We use standard assumptions for our simulations with \verb|PandExo| (without an associated noise floor), using the planetary and stellar parameters given in Table~\ref{tab:system-pars}, and assuming observational parameters corresponding to the NIRSpec PRISM observation of WASP-39~b. As we exclude a noise floor for this case, we see this as a more optimistic assumption on the associated transit depth error bars. We therefore refer to this as the ``optimistic'' noise case henceforth.  

\begin{figure}
					\centering
					\includegraphics[width=\hsize]{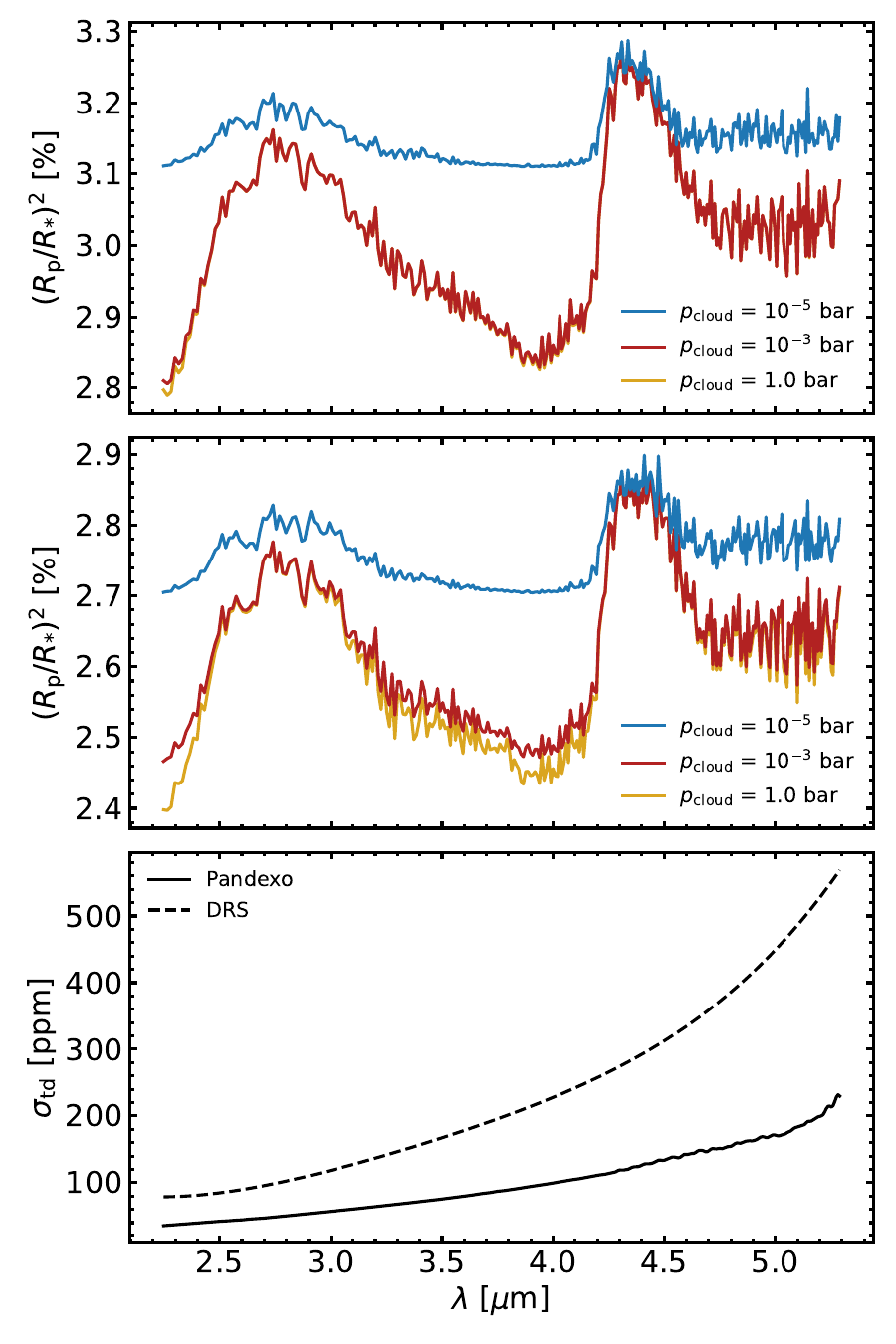}
					\caption{Synthetic forward model parameters that influence the generated transmission spectrum. 
					(Top) Transit depth (in \%) against wavelength (in \si{\micro\meter}) for the ``monotonic'' pressure-temperature profile. 
					(Middle) Same, but for the ``inverse'' pressure-temperature profile. 
					(Bottom) Associated transit depth uncertainty (in ppm) from a \texttt{PandExo} simulation (solid line), and for DRS case (dashed line).}
					\label{fig:synthetic-cases}
				\end{figure}

				We also include precision values associated with a \verb|Eureka!|-based reduction of the NIRSpec PRISM dataset of WASP-39~b. A more detailed description of this data reduction is given in Paper II, but the error bar estimation on the transit depth from this includes an additional point-scatter fitting parameter for the transit depth (e.g. \citetalias{jwsttransitingexoplanetcommunityearlyreleasescienceteam_2023}; \citealt{rustamkulov_2023}). We determine the transit depth precision values from this by fitting a fourth-order polynomial to the data reduction results to smooth over outliers and derive a more general error bar trend with respect to wavelength, referred to as ``data-reduction scaled'' (DRS) henceforth. This represents a more conservative approach to determine the error bars on the transit depth, resulting in values larger by a factor of approximately \num{2.3} when compared to the precision values from \verb|PandExo|. We include both error-bar cases to investigate and compare the influence of transit-depth S/N on the retrieval accuracy of the synthetic data. Both precision estimates considered in the construction of the synthetic transmission spectra are shown in the bottom panel of Fig.~\ref{fig:synthetic-cases}.

            \subsection{Retrieval performance metrics}\label{sec:metrics}

            We evaluate the performance of a retrieval with the chosen pressure-temperature characterisation based on two metrics: the Bayesian evidence, $E_\mathrm{M}$, associated with the model solution, and the accuracy of the retrieval with respect to the parameter ground-truths. 

                \subsubsection{Bayes' factor}

                We use $E_\mathrm{M}$ to calculate the Bayes' factor,
    			\begin{align}
    				B_{10} = \frac{E_1}{E_0},
    			\end{align}
    			where $E_0$ and $E_1$ refer to the evidence values associated with the underlying models $M_0$ and $M_1$, respectively. The Bayes' factor serves as a metric to quantify the evidence, based on the underlying data, in favour of a chosen model, $M_1$, with respect to a reference model, $M_0$ (or null-hypothesis). We evaluate the natural logarithm of the Bayes' factor, $\ln B_{10}$, equivalent to $\Delta \ln E$, following the prescription from \citet{kass_1995}, based on individual threshold values for $\ln B_{10}$ given in Table~\ref{tab:bf-evaluation}. 
       
                We note that the Bayes' factor is a relative metric comparing the fitting performance between different model assumptions, but does not make any statement about the absolute fitting performance of any single model. We omit an absolute metric from this work. The construction of the synthetic retrievals performed here tests for potentially inherent biases stemming from assumptions about the atmospheric p-T profile, but does not include any noise apart from assumed observational uncertainties. This implies that the fits performed here can reproduce the input data (meaning the transmission spectrum) to an indistinguishable degree, irrespective of the assumed multipoint p-T profile. We investigate how, for oversimplified p-T characterisations, this leads to compensatory deviations in other model parameters. We illustrate an overview of absolute fit-performances in Appendix~\ref{app:absolute-fit-performance}, but summarise this as follows -- the residuals of retrievals performed with any of the multipoint profiles, apart from the isothermal profile, are indistinguishable from each other. While we see a minor excess in residuals larger than the $1\sigma$-equivalent error-bar size in retrievals performed using the isothermal profile, 99\% of them fall below the associated $2\sigma$-value, and we see no individual residual with an absolute value larger than $3\sigma$.
    			
    			\begin{table}
    				\caption{Bayes' factor threshold values used to compare model performance.}
    				\label{tab:bf-evaluation}
    				\centering
    				\renewcommand{\arraystretch}{1.5}
    				\begin{tabular}{cc}
    					\hline\hline
    					Value & Description \\
    					\hline
    					$1 < \ln B_{10} < 3$	& positive evidence for model 1 \\
    					$3 < \ln B_{10} < 5$	& strong evidence for model 1 \\
    					$\ln B_{10} > 5$		& very strong evidence for model 1 \\
    					\hline
    				\end{tabular}
    				\tablefoot{
    					The threshold values of $\ln B_{10}$ correspond the formalism suggested by \citet{kass_1995}. We exclude $\vert \ln B_{10} \vert < 1.0$, which represents an inconclusive statement about model preference. Negative values of $\ln B_{10}$ have inverse meanings, representing evidence in favour of model 0, the null-hypothesis.
    				}
    			     \end{table}

                \subsubsection{Retrieval accuracy}\label{sssec:parameter-accuracy}

                    Additionally, we evaluate the performance of each retrieval against the known input parameter values of the underlying synthetic spectra. To evaluate the accuracy of the retrieved parameters, we calculate the distance between the estimated parameter, $\theta_\mathrm{p,est}$, and the known true value, $\theta_\mathrm{p,GT}$,
                    \begin{align}
        				\Delta \theta_\mathrm{p} = \theta_\mathrm{p,est} - \theta_\mathrm{p,GT},
        			\end{align}
                    for the planetary reference radius, $R_\mathrm{p}$, the VMRs given in Table~\ref{tab:synth-pars}, and the cloud-top pressure, $p_\mathrm{cloud}$, corresponding to the individual synthetic forward model cases described in Sect.~\ref{sec:mock-data}. This value will be positive if the retrieved parameter value overestimates the underlying true value, and negative if the underlying true values are underestimated. 
                    
                    We scale $\Delta \theta_\mathrm{p}$ using a factor representing the attributed uncertainty in the estimated parameter value, employing two different prescriptions. Firstly, we calculate a centred credible interval (CCI) containing a $3\sigma$-equivalent of the sample-parameter population. For this, we determine the centre of the credible interval to be the median of the posterior distribution (or the 50\%-quantile), and the edges to be defined by the quantile values encompassing 99.7\% of the posterior distribution. Additionally, we use the $1\sigma$ value given by \texttt{TauREx} as a scale factor, which is computed as the CCI encompassing 68\% of the posterior distribution. An example of the parameter estimation evaluation is shown in Fig.~\ref{fig:parameter-accuarcy-example-drs} for the case of a synthetic transmission spectrum generated with an underlying ``inverse'' p-T profile, a cloud-top pressure of $10^{-5}$ \si{\bar}, and DRS transit-depth error bars. The asymmetry of the upper and lower boundaries around the median value of the retrieved parameter derives from the underlying non-gaussianity of the marginalised posterior distribution. We provide a more detailed view of all accuracy values associated with individual parameters in the supplementary repository for this work\footnote{\url{https://doi.org/10.5281/zenodo.10497342}}. In the case of the $3\sigma$-equivalent CCI, we evaluate the retrieval accuracy on a binary scale, with the true parameter value either being within the credible interval, or not. For the $1\sigma$ intervals provided by \texttt{TauREx}, we compute the scaled parameter distance outright.

    \section{Results and discussion}\label{sec:results}

        To quantify the performance of different pressure-temperature prescriptions on the recovery of atmospheric characteristics, we perform a homogenised set of retrievals on the sample of synthetic transmission spectra, using a variety of multipoint pressure-temperature profiles. As a reference case, we perform retrievals using a free p-T profile with eleven pressure nodes, which represents a self-retrieval on the constructed input cases described in Sect.~\ref{ssec:synthetic-pt}. Firstly, we compare the performance of this to retrievals that utilise an isothermal prescription, equivalent to fitting one temperature parameter. We then iteratively increase the number of pressure nodes, performing retrievals using a two-point profile (which represents the assumption that a temperature gradient in the atmosphere can be retrieved), a four-point profile (allowing for the retrieval of a potential temperature inversion point), as well as a six-point and eight-point profile (to investigate the potential of recovering an even more detailed p-T structure than with the two- or four-point profile).

        \subsection{Retrieving synthetic spectra}
        
\begin{figure*}
            \centering
            \includegraphics[width=\hsize]{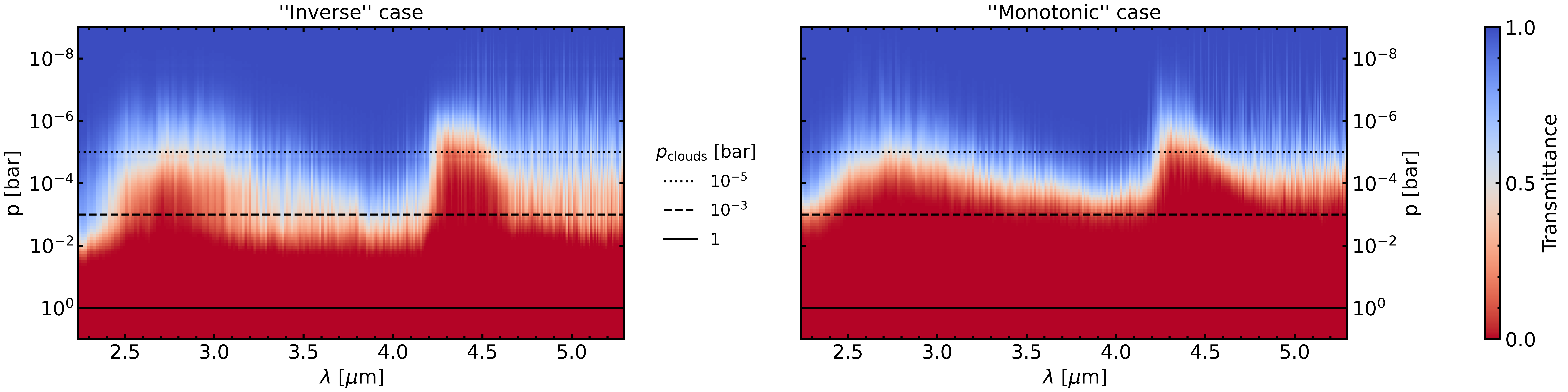}
            \caption{Transmittance for the sample of synthetic spectra generated with the ``monotonic'' (left panel) and ``inverse'' (right panel) input pressure-temperature profile. Both panels show wavelength (in \si{\micro\meter} on the x-axis, and pressure (in \si{\bar}) on the y-axis. For illustrative purposes, the transmittance in these plots is calculated without the cloud-deck cases. The cloud-top pressure levels, $p_\mathrm{cloud}$ considered in this work (as described in Sect.~\ref{ssec:cloud-cases}) are marked with a solid ($p_\mathrm{cloud} = 1\,\si{\bar}$), dashed ($p_\mathrm{cloud} = 10^{-3}\,\si{\bar}$), and dotted ($p_\mathrm{cloud} = 10^{-5}\,\si{\bar}$) black line.}
            \label{fig:transmittance-plots}
        \end{figure*}

        In all cases, we use \verb|TauREx| with the general setup described in Sect.~\ref{sec:methods}, and use the same forward model setup as in the construction of the transmission spectra, i.e. the pressure domain, assumption of homogeneous VMRs, and atmospheric opacity sources. For all cases, we retrieve the planetary radius at the bottom of the atmospheric pressure regime, $R_\mathrm{p}$, molecular VMRs, $\log_{10}(X_\mathrm{VMR})$, and cloud-top pressure, $\log_{10}(p_\mathrm{cloud})$, representing seven free parameters shared by all models. In addition, for each model we fit the free parameters associated with the p-T prescription. In the case of an isothermal profile, this means retrieving the temperature parameter $T_\mathrm{iso}$, while in the case of the multipoint profiles, we retrieve the temperature values, $T_\mathrm{n}$, at fixed pressure nodes, $p_\mathrm{n}$, within the atmosphere. These pressure nodes are set to the top (TOA) and bottom (BOA) of the atmosphere for the first and last node of each case, respectively. The remaining nodes are distributed close to log-uniform within the pressure domain -- we separate them as equally as possible between the top and bottom of the atmospheric pressure domain, but make an informed choice to slightly cluster them around the expected probed atmospheric pressure region. This is illustrated in Fig.~\ref{fig:transmittance-plots}, which shows the transmittance of the atmospheric forward model for both the spectra generated with the ``inverse'', as well as with the ``monotonic'' pressure-temperature profile, calculated without the respective cloud-deck cases. The probed atmospheric pressure regime, excluding the cloud deck, extends approximately between \SI{1}{\milli\bar} and \SI{10}{\micro\bar}. The explicit values of $\log_{10}(p_\mathrm{n} \, \mathrm{[bar]})$ for these multipoint profiles are
        \begin{itemize}
            \item $\left\{ 1, -3, -7, -9 \right\}$ for the four-point profile,
            \item $\left\{ 1, -1, -3, -5, -7, -9 \right\}$ for the six-point profile,
            \item $\left\{ 1, -1, -2, -3, -4, -5, -7, -9 \right\}$ for the eight-profile.
        \end{itemize}

        We note that the pressure-nodes for the self-retrieval performed with an eleven-point p-T profile are the same ones shown in Fig.~\ref{fig:pT-GT}. In total, this results in a number of free parameters between eight (isothermal case) and eighteen (eleven-point pressure-temperature profile). We list these retrieved parameters, and their associated prior distributions, in Table~\ref{tab:synth-priors}.

        \begin{table}
            \caption{Priors used in atmospheric retrieval of synthetic spectra}
            \label{tab:synth-priors}
            \centering
            \renewcommand{\arraystretch}{1.5}
            \begin{tabular}{cccc}
                \hline\hline
                Parameter & Prior Type & Range & Assoc. Unit \\
                \hline
                $R_\mathrm{p}$		& Uniform		& [\num{1.0};\,\num{1.5}]	& \si{\jupiterradius} \\
                $T_\mathrm{i}$		& Uniform		& [\num{300};\,\num{3000}]	& \si{\kelvin} \\
                $X_\mathrm{VMR}$ 	& LogUniform	& [\num{-12};\,\num{-0.1}]	& - \\
                $p_\mathrm{cloud}$	& LogUniform	& [\num{1};\,\num{-9}]		& \si{\bar} \\
                \hline
            \end{tabular}
            \tablefoot{
                $T_\mathrm{i}$ refers to temperature points within multipoint profiles, where in the case of an isothermal profile, $T_\mathrm{i} = T_\mathrm{iso}$, and in the case of an n-point profile, $T_0 = T_\mathrm{BOA}$ and $T_\mathrm{n} = T_\mathrm{TOA}$. The VMRs for atmospheric trace gases (\ce{H2O}, \ce{CO}, \ce{CO2}, \ce{H2S}, \ce{CH4}) are constant throughout the atmosphere.
            }
        \end{table}		

\begin{figure*}
            \centering
            \includegraphics[width=\hsize]{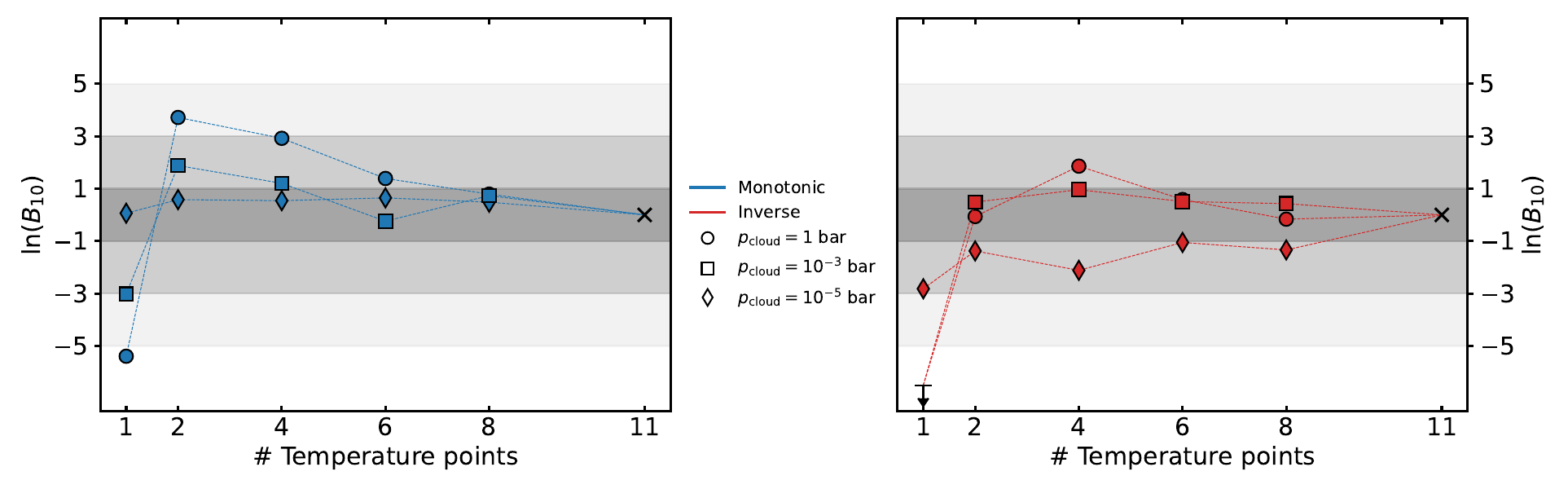}
            \caption{Bayesian evidence comparison for retrievals on all cases of the synthetic dataset associated with the DRS noise case, split into the synthetic spectra generated using the ``monotonic'' input p-T profile (left), and the ``inverse'' input p-T profile (right). Both panels show the number of retrieved temperature points (corresponding the implemented multipoint profile) on the x-axis, and the model-associated Bayes' factor on the y-axis. Data points are separated into spectra generated with high ($p_\mathrm{cloud} = \SI{1}{\bar}$, circles), medium ($p_\mathrm{cloud} = 10^{-3}\,\si{\bar}$, squares), and low ($p_\mathrm{cloud} = 10^{-5}\,\si{\bar}$, diamonds) cloud-top pressure input values. The grey shaded are denotes the Bayes' factor threshold given in Table~\ref{tab:bf-evaluation}. The black cross denotes the reference model in all cases (eleven-point p-T profile), and Bayes' factor values fulfilling $\vert\ln B_{10}\vert > 6.0$ are denoted by a black arrow.}
            \label{fig:bf-drs}
        \end{figure*}

        We illustrate the evaluation of models based on the Bayes' factor, $\ln B_{10}$, in Fig.~\ref{fig:bf-drs}, showing the Bayes' factor in reference to the self-retrieval using the eleven-point profile, as a function of the number of the temperature points in each model (exemplified on the DRS noise case). In the majority of cases, the isothermal model is strongly disfavoured against all other multipoint models, illustrating the fact that the isothermal prescription of the p-T profile is not capable of accurately capturing the complexity of either of the underlying pressure-temperature profiles shown in Fig.~\ref{fig:pT-GT}. The exception to this are the synthetic transmission spectra generated with a low-pressure cloud top ($p_\mathrm{cloud} = 10^{-5} \, \si{\bar}$). In these cases, the Bayes' factor does not sufficiently distinguish between any of the models, owing to the fact that a flat-opacity cloud-deck at pressure-levels this low obscures significant fractions of the molecular signatures, masking the impact of the p-T profile on the molecular contributions. This can also be seen in Fig.~\ref{fig:transmittance-plots}, where the low-pressure cloud deck as a flat-opacity step function obscures a significant fraction of the wavelength-dependent transmittance at higher pressures, effectively masking the probed pressure regime.
        
        Apart from the disfavoured performance of the isothermal p-T profile, the behaviour of the Bayes' factor is clearly separated between the cases of an underlying ``monotonic'' p-T profile (left panel of Fig.~\ref{fig:bf-drs}), and an underlying ``inverse'' p-T profile (right panel of Fig.~\ref{fig:bf-drs}). In the former case, the two- and four-point profiles are strongly to moderately preferred, while the Bayes' factors for profiles with a higher degree of complexity converge toward zero. However, in the ``inverse'' case, the two-point profile is no longer preferred against the self-retrieval, while the four-point profile still sees moderate preference in the high-pressure cloud case. The model performance evaluation for the ``optimistic'' noise case, shown in Fig.~\ref{fig:bf-pandexo}, provides conclusions in line with the DRS case, with the distinction that the smaller transit-depth error bars associated with the ``optimistic'' case lead to more distinctly differentiated Bayes' factors for each model. This is in line with the expectation that the increased S/N of the transmission spectrum in these cases provides a higher information content. The ``optimistic'' noise case provides a clearer preference for the two-point and four-point profile in the ``monotonic'' input p-T case, and a clearer Bayes' factor peak for the four-point profile used in the retrievals on spectra generated with the ``inverse'' p-T profile.
        
        From the evaluation of $\ln B_{10}$, we can conclude that the isothermal profile should be disregarded compared to more complex p-T characterisation when retrieving transmission spectra of a quality in line with expected JWST observations. This is in agreement with the conclusions presented in \citet{rocchetto_2016}, who showed that in retrievals of transmission spectra modelled for expected JWST observations of hot Jupiters, assuming an isothermal p-T profile leads to a significant bias in the retrieved atmospheric VMRs.
			
\begin{figure*}
            \centering
            \includegraphics[width=\hsize]{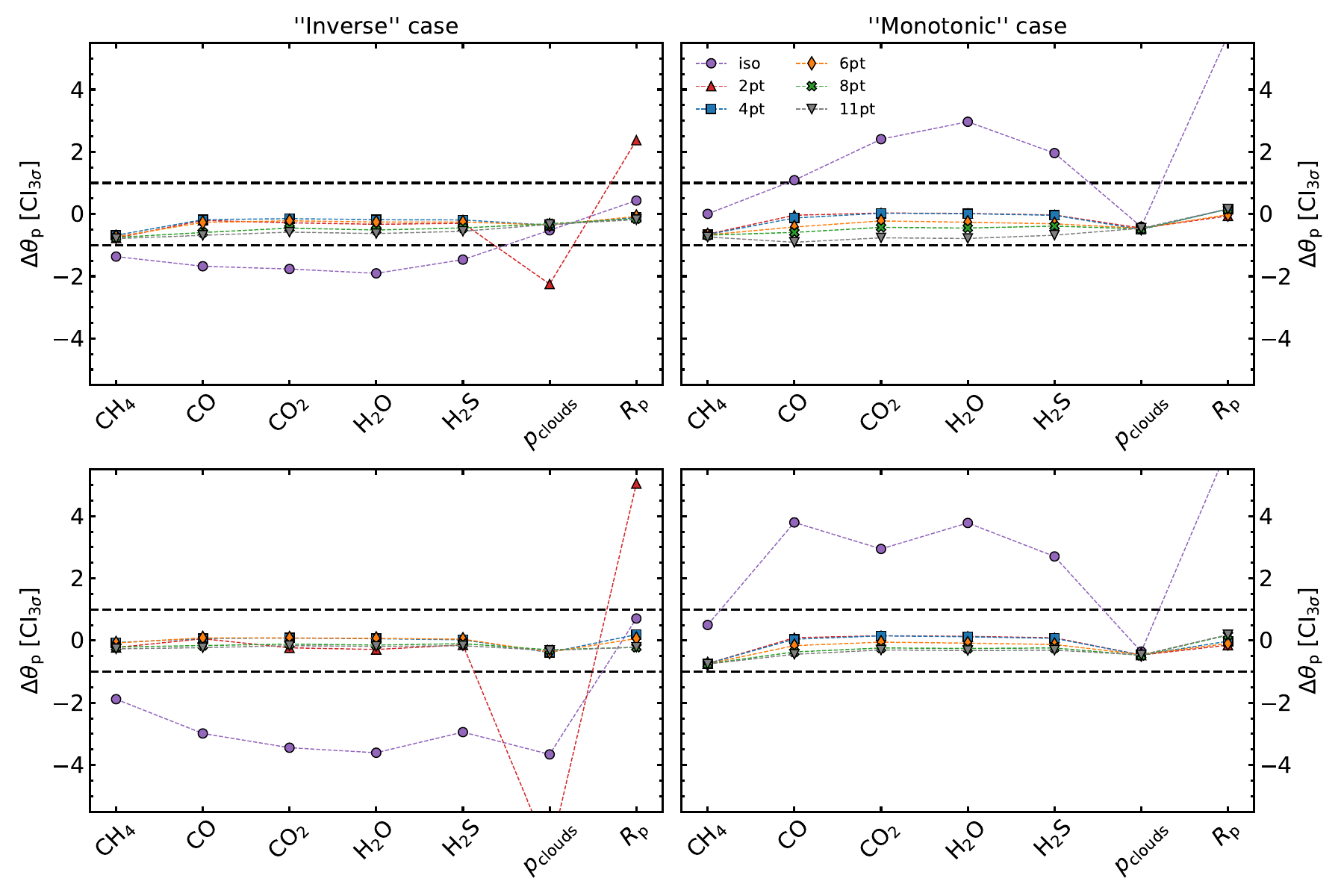}
            \caption{Accuracy of retrieved parameters, showing individual parameters on the x-axis, and accuracy calculated based on the 3$\sigma$-equivalent centred credible interval on the y-axis. The top two panels represent retrievals performed on a synthetic spectrum generated under the DRS noise assumption, while the bottom two panels show the same for the ``optimistic'' noise case. (Left) Results for retrievals performed on a synthetic spectrum generated with the ``inverse'' input p-T profile. (Right) Same, but for a synthetic spectrum generated with the ``monotonic'' p-T profile. In all panels, the reported accuracy values are colour-coded by the p-T profile used in the retrieval. The black dashed lines denote the boundaries of the $3\sigma$ CCI, and associated accuracy values of $\vert \Delta\theta_\mathrm{p} \vert > 5.0$ are cut-off for visualisation purposes.}
            \label{fig:retrieval-accuracy-example}
        \end{figure*}

        The accuracy of the retrieved parameters corroborates the preference for a more complex prescription of the pressure-temperature profile. Retrievals performed using an isothermal profile consistently lead to biased parameters, in several cases with misleadingly high precision. We illustrate this in Fig.~\ref{fig:retrieval-accuracy-example}, which shows the retrieval accuracy evaluated using the 3$\sigma$-equivalent CCI for synthetic transmission spectra containing a high-pressure cloud-top, for both assumed noise cases. In all cases, performing retrievals with the isothermal prescription fails to retrieve the VMRs of the main atmospheric molecular species. The exception to this, in the ``monotonic'' case, is \ce{CH4}, which, due to the low underlying abundance value and lack of distinct absorption signatures, has significantly broader posterior distributions. Following from this, the associated parameter estimation uncertainty values are significantly larger, which more readily encompass the underlying true value. In the case of the other retrieved molecular VMRs, the value of $\Delta\theta_\mathrm{p}$ for the isothermal case is centred at around twice the size of the 3$\sigma$ CCI.

        When comparing the accuracy of the retrieved parameters between retrievals performed on synthetic spectra generated under the DRS noise assumption (as exemplified by the top two panels in Fig.~\ref{fig:retrieval-accuracy-example}), and on spectra generated using the ``optimistic'' noise assumption (as exemplified by the bottom two panels in Fig.~\ref{fig:retrieval-accuracy-example}), it is also readily apparent that in the latter case, the values of $\Delta\theta_\mathrm{p}$ are, on the scale of the 3$\sigma$ CCI, larger than for the equivalent DRS case. This is influenced by the higher reported retrieval precision, reflected in a smaller centred credible interval. An illustration of this can be seen in Appendix~\ref{sec:retrieval-accuracy-expl}, which shows a comparison between credible intervals of the retrieved parameters for all multipoint profile cases. We provide a complete overview of the accuracy evaluation for all parameters and cases in Figs.~\ref{fig:parameter-accuracy-drs-inv} - \ref{fig:parameter-accuracy-pandexo-mon}.

        As suggested by the Bayes' factor analysis for model preference, distinguishing between favoured multipoint characterisation for the pressure-temperature profile is less clear. In the case of of ``monotonic'' underlying p-T structure, we find equal performance for all tested multipoint profiles, regardless of the accuracy scale factor used in evaluating them. The pressure-temperature structure of this profile, shown in Fig.~\ref{fig:pT-GT}, suggests that this can be expected, as a simple atmospheric temperature gradient, produced by the two-point profile, can be judged as sufficient to reproduce the overall structure of the atmospheric p-T profile. Considering either approach in evaluating the retrieval accuracy shows the equivalently successful performance of all multipoint profiles in retrieving the correct true parameter values. Based on the Bayes' factor evaluations in Fig.~\ref{fig:bf-drs} and Fig.~\ref{fig:bf-pandexo}, the two-point and four-point profiles are the most favoured ones, where the two-point profile could be preferred overall due to its smaller amount of free parameters.
			
        This picture changes in the case of the ``inverse'' input p-T profile. Comparing the structure of this underlying profile (see also Fig.~\ref{fig:pT-GT}) with the expected p-T structure in retrieving a two-point profile, it is apparent that the two-point profile does not provide enough complexity to capture the temperature inversion point. This leads here to wrongly retrieving the planetary reference radius and cloud-top pressure in several cases, one of which is illustrated in the left panels of Fig.~\ref{fig:retrieval-accuracy-example}. Distinguishing the performance of retrievals using a pressure-temperature profile with more than two points based on the accuracy alone proves to be more inconclusive. As would be expected for models with an increasing number of free parameters, we see no more distinguishable difference in the retrieved parameter accuracy for any of these cases.

\begin{figure*}
            \centering
            \includegraphics[width=\hsize]{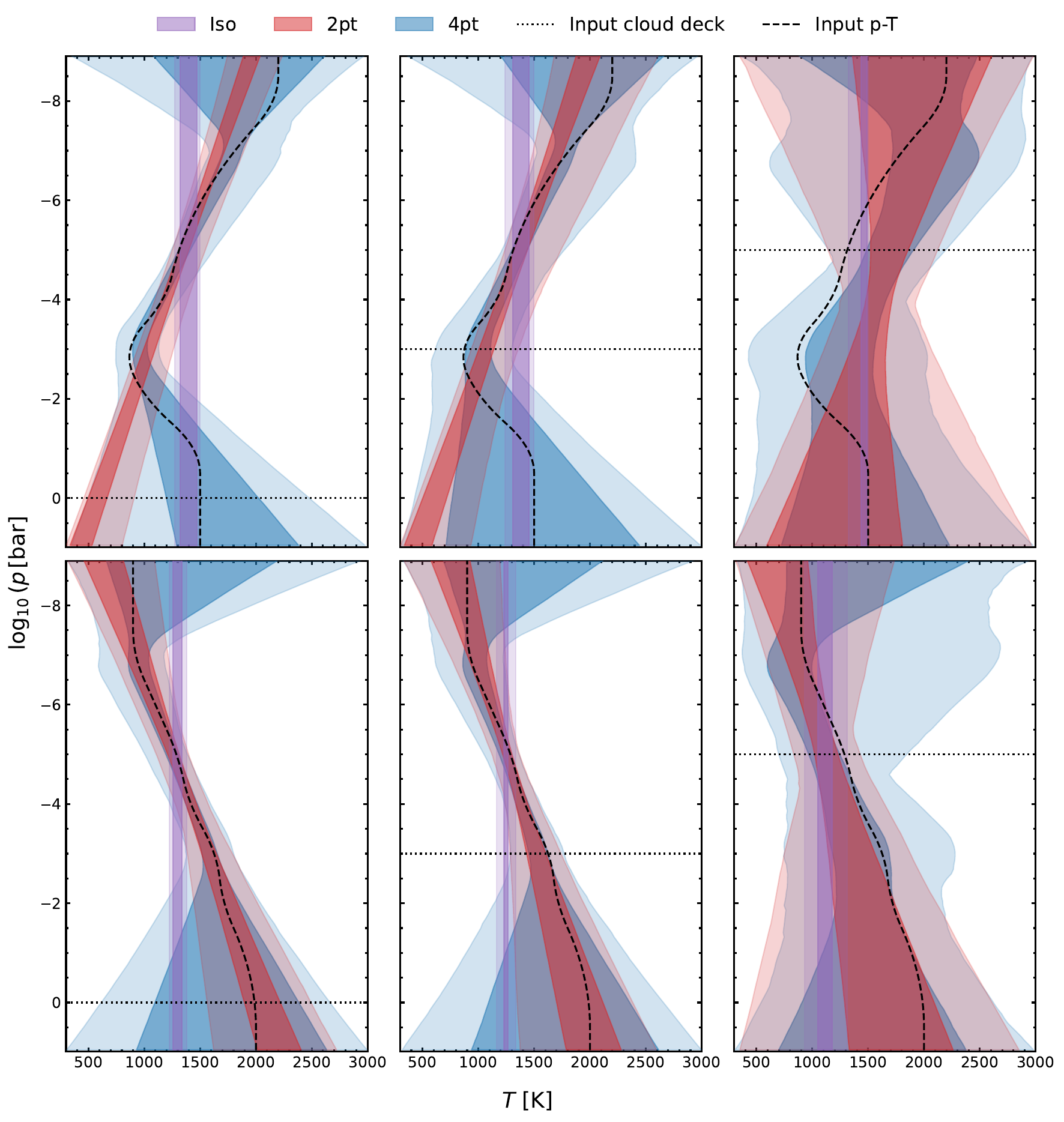}
            \caption{
                Comparison between retrieved pressure-temperature profiles and corresponding ground truths, showing temperature (in \si{\kelvin}) on the x-axis, and pressure (in \si{\bar}) on the y-axis, for synthetic cases using the error bar size scaled from data-reduction results. The first three panels show the ``inverse'' p-T profile as the ground truth, and are ordered by increasing cloud-top pressure, $\log_{10}(p_\mathrm{cloud} \, [\mathrm{bar}]) = \{ 0, -3, -5 \}$. The second three panels show the same, but for the ``monotonic'' p-T profile as the ground truth.
            }
            \label{fig:accuracy-pt-profiles-drs}
        \end{figure*}

        The connection between the preference of different multipoint profiles to the underlying pressure-temperature structure of the input spectrum is further illustrated in Fig.~\ref{fig:accuracy-pt-profiles-drs}, which shows the retrieved p-T profiles of the isothermal, two-point, and four-point assumption in the context of the DRS noise case. With an underlying ``inverse'' p-T profile, a multipoint profile with more than two pressure nodes is necessary to capture the input p-T structure. A two-point profile, which can only produce a temperature gradient under the assumptions taken here, cannot sufficiently reproduce the inversion point of this profile. On the other hand, in the case of the ``monotonic'' input case, it becomes quite apparent that a two-point profile is already sufficient in capturing the, at first order, gradient-like structure of the input p-T profile. In either case, the isothermal profile is, by its construction, unable to reproduce any vertical temperature structure in the atmosphere, and fails to accurately retrieve a majority of the main atmospheric characteristics. This can also be seen in the bottom panels of Fig.~\ref{fig:accuracy-pt-profiles-drs}, which shows that, apart from the isothermal profile, each of the retrieved multipoint profiles is in good agreement with the underlying p-T structure used as the input for the corresponding synthetic transmission spectrum. We show the retrieved pressure-temperature profiles for the synthetic cases associated with the ``optimistic'' noise case in Appendix~\ref{sec:additional-pt-profile}.

        We note that in all retrievals run under an isothermal assumption, a distinct difference between the accuracy of the retrieved atmospheric parameters is visible when comparing the input cases using the ``inverse'' p-T profile (Figs.~\ref{fig:parameter-accuracy-drs-inv} and \ref{fig:parameter-accuracy-pandexo-inv}), and the ``monotonic'' p-T profile (Figs.~\ref{fig:parameter-accuracy-drs-norm} and \ref{fig:parameter-accuracy-pandexo-mon}). With an underlying ``inverse'' profile, the value of $\Delta\theta_\mathrm{p}$ for all molecular VMRs is consistently negative, implying that the parameters are underestimated during the retrieval process. In the cases with an underlying ``monotonic'' p-T profile, these values are consistently positive, implying that the parameters are overestimated. As shown in Fig.~\ref{fig:transmittance-plots}, the probed pressure regime in both cases extends approximately between \SI{1}{\milli\bar} and \SI{10}{\micro\bar}. Comparing this to the retrieved p-T profile for the isothermal case (Figs.~\ref{fig:accuracy-pt-profiles-drs} and \ref{fig:accuracy-pt-profiles-pandexo}), it is apparent that in the ``monotonic'' case, the temperature profile is lower than the underlying ground truth, while in the ``inverse'' case, the opposite is true. This anti-correlation between the biases in temperature and atmospheric VMRs is attributable to the influence of the atmospheric scale-height in the transit geometry. If the temperature is overestimated, the scale-height of the atmosphere increases. Subsequently, this also increases the path-length within the atmospheric layer. In the calculation of the optical depth, the increased path-length is compensated by lowering the chemical abundances (see Equation 5 in \citealt{al-refaie_2021}, represented by the column-density). If the temperature is underestimated, the opposite effect takes place.

		\subsection{Outlook}\label{sub-sec:limitations}

			We note that the synthetic retrieval investigation described above only represents a preliminary analysis of evaluating the performance of an isothermal prescription against that of more complex multipoint profiles. In this work, we have chosen the underlying pressure-temperature profiles with arbitrary complexity to cover a simpler, positive temperature lapse rate, and a more complex structure including a temperature inversion point, respectively. Similarly, we have modelled the chemical structure of the mock-atmosphere after previously described results for the hot Jupiter WASP-39~b (\citetalias{jwsttransitingexoplanetcommunityearlyreleasescienceteam_2023}), and did not incorporate physics-informed choices of the atmospheric chemical structure based on the chosen p-T profile. Additionally, in this work we have only looked at the case of a primordial, \ce{H2 / He}-dominated atmosphere, which, due to their low mean molecular weight, are prime targets for detailed atmospheric characterisations efforts. A more comprehensive study aiming to address this question for a broader variety of p-T structure and planetary characteristics would benefit the overall conclusions drawn here.

            In this work, we have constructed an input transmission spectrum using VMRs that are constant with altitude. This guarantees that, when performing atmospheric retrievals under the same assumption of constant profiles, we do not introduce additional biases in oversimplifying the vertical VMR structure during the retrieval. However, the assumption about constant VMRs might be considered an oversimplification, along the same lines as an isothermal p-T profile. As shown in Fig.~\ref{fig:retrieval-accuracy-example}, there is a degeneracy between the retrieved temperature and molecular VMR values. Through their influence on the atmospheric scale-height, the biases in temperature and VMR values are anti-correlated. This implies that an oversimplified chemical profile used for atmospheric retrievals could induce a similar bias to the retrieved pressure-temperature structure. \citet{changeat_2019} have shown that, in the context of simulated observations of JWST and Ariel, using constant chemical profiles when the underlying true chemical profile is more complex introduces this degenerate bias in retrieving an isothermal p-T profile. While a test of this manner is beyond the scope of this work, additional investigations to catalogue the effects of these sources of biases in atmospheric retrievals will help inform best practices for characterisation of exoplanet atmospheres in a consistent manner.

			We also point out that, in the case of multipoint pressure-temperature profiles, the location of the pressure nodes will influence the performance of the associated retrieval, especially with respect to reproducing ground-truth p-T profiles, as is the case here. Fixing the pressure nodes inherently includes the possibility that a significant ``feature'' of the underlying p-T structure (such as the inversion point of the ``inverse'' profile tested here) is missed. This could be counteracted by including the pressure-nodes of the profile as free parameters as well. However, this will in effect double the number of free parameters associated with any multipoint profile, which will significantly increase the computational requirements of running such retrievals \citep{changeat_2021}. Another approach to mitigate this problem could be provided by determining a pressure-region of maximum contribution function to the optical depth (or transmittance) of the atmosphere, and clustering any pressure-nodes around this relevant region. \citet{waldmann_2015} described an approach of this manner for the retrieval of exoplanetary emission spectra, performing a multi-stage approach to dynamically adjust the complexity of the retrieved p-T profile. A detailed study of the application of this to transmission spectra is outside the scope of this work, but would provide additional insight about best practices in determining a free p-T structure of exoplanet atmospheres from transmission spectroscopic observations.

            Applying atmospheric retrievals to real data sets adds observational noise characteristics as a potential additional source of bias in the atmospheric characterisation process. A recent, in-depth analysis of the information content of the JWST transit observations of WASP-39~b by \citet{lueber_2024} has explored the question of p-T complexity in the case of real data. The authors found -- by performing a Bayes' factor analysis on atmospheric retrievals of the individual near-infrared transmission spectra taken by JWST -- preference for isothermal prescriptions in several cases. Based on the results shown here, we advocate to implement p-T prescriptions of at least the complexity of a two-point profile. If atmospheric retrievals using more complex p-T prescriptions show disagreement in the retrieved chemical profiles when compared to retrievals performed using an isothermal prescription, it would indicate an unresolved underlying bias. We present one possible source of this, although other sources of model biases could also impact transmission results. For instance, oversimplified chemical profiles \citep{changeat_2019}, 3D effects \citep[e.g.][]{espinoza_2021,pluriel_2022}, and poorly constrained planetary masses \citep[e.g.][]{changeat_2020-1,dimaio_2023} have already been investigated in the past. To quantify the importance of such biases, retrieval results should systematically be evaluated against complex prescriptions in controlled scenarios.

    \section{Conclusions}\label{sec:conclusion}

        We investigated the recoverable complexity of pressure-temperature profiles associated with transmission spectra observed with JWST, based on synthetic transmission spectra modelled after the hot Jupiter WASP-39~b. We investigated several scenarios, including two different underlying pressure-temperature structures, several different cloud-top pressure values, and two different assumptions about the associated transit depth uncertainty, to cover a large range of possible input transmission spectra. We tested the performance of atmospheric retrievals using an isothermal, as well as multipoint pressure-temperature profiles with two, four, six, and eight fixed pressure nodes. We evaluated the performance of these individual retrievals based on two metrics -- the Bayes' factor, and the accuracy of the retrieved parameters in comparison to the known true values. The Bayes' factor was evaluated in reference to retrievals performed with an eleven-point pressure-temperature prescription, which represents a self-retrieval of the input p-T profile.
        
        Based on these two metrics, we find that the isothermal profile is an insufficient approximation for atmospheric retrievals of transmission spectra with JWST-level expected precision. We find that the Bayes' factor analysis consistently rejects retrievals run with the isothermal prescription, and we additionally see that these retrievals consistently lead to wrongly recovered atmospheric parameters. In comparing the performance of the individual multipoint profiles more complex than the isothermal case, the Bayes' factor analysis is able to distinguish preferred profiles based on the underlying input p-T case. Both the two- and four-point profiles show moderate to strong preference when the underlying pressure-temperature profile shows a purely positive lapse rate. In the case of an underlying p-T profile showing an inversion point, we find that the four-point prescription is the only preferred profile. Compared to this, more complex multipoint profiles are not preferred in any cases, most likely owing to the fact that their additional free parameters penalise the Bayesian evidence evaluation. In comparing the accuracy of the retrieved parameters, we find sporadically wrong values for the two-point profile, but see no clear distinguishing factor for the more complex prescriptions.
        
        This suggests that, among the p-T characterisations investigated in this work, the four-point profile is the least complex model that correctly retrieves the known input parameters in all cases, and leaves flexibility in the retrieval to capture potentially more complex pressure-temperature structures. While a fine-grained selection process of an optimal multipoint p-T profile based on the expected nature of the transmission spectrum could still be improved in future work, we recommend not to use an isothermal prescription, as it has been shown to wrongly bias the retrieved atmospheric VMRs. This bias also depends on the underlying p-T structure, being positive in the case of a positive temperature lapse rate, but negative in the presence of an temperature inversion point around the probed atmospheric pressure regime.


	\begin{acknowledgements}

		We thank the anonymous referee for their helpful comments and suggestions. S. Schleich extends his gratitude to G. Van Looveren, N. Pawellek, and M. Voyer for insightful discussions on the representation of the results. This project was funded by the FGGA Emerging Field Grant 2021. We acknowledge financial support by the University of Vienna and Österreichische Forschungsgemeinschaft (ÖFG). Parts of this work were achieved using the Vienna Scientific Cluster (VSC). \\
        \ \\
        \textit{Data and software statement.} We provide a collection of the data products described in this work, as well as ancillary data, at this repository: \url{https://doi.org/10.5281/zenodo.13737625}. We gratefully acknowledge the use of open-source libraries for the Python programming language: \textbf{matplotlib} \citep{hunter_2007}, \textbf{numpy} \citep{harris_2020}, and \textbf{pandas} \citep{mckinney_2010}.
		
	\end{acknowledgements}

	\bibliography{references.bib}
	\bibliographystyle{aa}

    \begin{appendix}

		\section{Atmospheric opacity data}

			Table~\ref{tab:atm-opacity-data} shows an overview of the opacity data and associated references used within the retrievals conducted in this work.
			
			\begin{table}[!ht]
				\caption{References for atmospheric opacity data}
				\label{tab:atm-opacity-data}
				\centering
				\renewcommand{\arraystretch}{1.5}
				\begin{tabular}{cc}
					\hline\hline
					Opacity source & Reference \\
					\hline
					\multicolumn{2}{c}{Molecules} \\
					\hline
                    \ce{CH4}	& \citet{yurchenko_2024} \\
					\ce{CO}		& \citet{li_2015} \\
					\ce{CO2}	& \citet{yurchenko_2020} \\
					\ce{H2O}	& \citet{polyansky_2018} \\
					\ce{H2S}	& \citet{azzam_2016} \\
					\hline
					\multicolumn{2}{c}{Collision-induced absorption} \\
					\hline
					\ce{H2}-\ce{H2}	& \citet{abel_2011}, \citet{fletcher_2018} \\
					\ce{H2}-\ce{He}	& \citet{abel_2012}\\
				\end{tabular}
			\end{table}


        \FloatBarrier
        \section{Absolute retrieval fit performance}\label{app:absolute-fit-performance}

\begin{figure}
                \centering
                \includegraphics[width=.75\hsize]{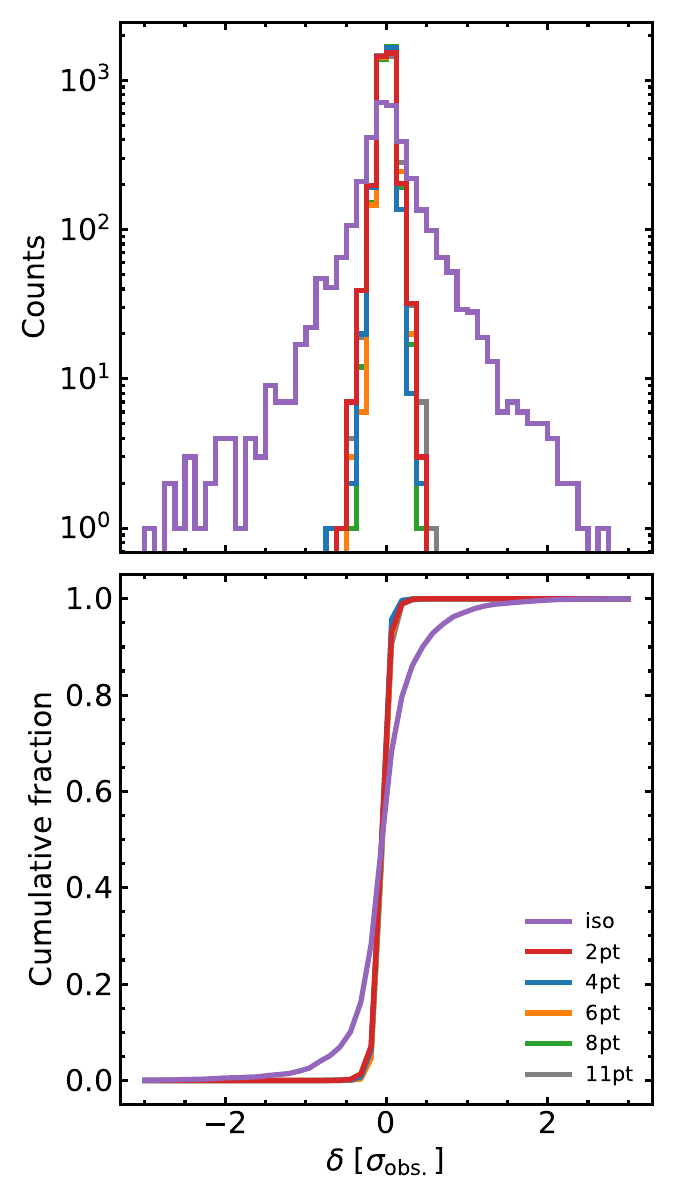}
                \caption{Distributions of retrieval residuals (scaled to the observational error bar, $\sigma_\mathrm{obs.}$), counted in bins with a width of 0.25$\sigma_\mathrm{obs}$. (Top) Total number of counts per bin. (Bottom) Cumulative fraction of residuals. Both panels show the collected distribution of residuals from retrievals applied to each of the twelve synthetic transmission spectra, but counted separately for each pressure-temperature prescription.}
                \label{fig:total-residual-histogramme}
            \end{figure}

            In evaluating the atmospheric retrievals of the synthetic transmission spectra described in Sect.~\ref{sec:mock-data}, we utilise the Bayes' factor, $\ln/B_{10}$, as a relative metric to compare the performance of individual model. As the Bayes' factor is a relative metric, comparing the evidence value, $\ln(E)$, associated with each model, it makes no comment about the absolute fit performance for each run. To review to total performance of each model, differentiated only by the characterisation of the p-T profile, we investigate the total distribution of individual fit residuals, $\delta_i$, scaled to the associated data point error bars,

            \begin{align}
                \delta_i = \frac{X_i - O_i}{\sigma_i},
            \end{align}

            where for each data point of the synthetic transmission spectrum, $O_\mathrm{i}$, we calculate the residual associated with the best-fit model point, $X_\mathrm{i}$, and scale it by the error-bar size, $\sigma_\mathrm{i}$, associated with the data point. Figure~\ref{fig:total-residual-histogramme} (top panel) illustrates this distribution for each pressure-temperature profile used in our retrieval runs, where the residuals for each of the twelve synthetic input spectra are combined. The distributions of residuals for the two-, four-, six-, eight-, and eleven-point profiles are indistinguishable from each other. In comparison to that, the residuals for all retrievals using the isothermal profile show a broader wings towards larger values. However, we still find that, in a cumulative count of all residual values from isothermal retrievals, 99.2\% fall within $\vert\delta\vert < 2.0$, and 95\% fall within $\vert\delta\vert < 1.0$ (bottom panel of Fig.~\ref{fig:total-residual-histogramme}). This is attributable to the way in which the synthetic spectra in our work are set up -- Our investigation aims to find inherent biases stemming from the utilisation of different p-T profile complexities, which is why we omit this scatter process to not introduce uncontrollable biases \citep{feng_2018, changeat_2019}. As we do not add any scatter do the data points in the construction of the transmission spectra, we enable the fitting process to very closely reproduce the shape of the underlying cross-sectional data used to generate the spectra. Based on this, we deem it not necessary to discuss the absolute performance of each fitting process in the retrievals we conduct, and restrict our analysis to the relative model comparison through the Bayes' factor, as well as the accuracy evaluation in the known underlying input parameters of each synthetic transmission spectrum.
            
        \clearpage\newpage
        \FloatBarrier
        \begin{onecolumn}
        \section{Additional Bayes' factor analysis}

			In addition to the results of synthetic retrievals on the mock atmospheric transmission spectra created with error bars scaled by data reduction results, Fig.~\ref{fig:bf-pandexo} shows the same analysis on spectra created with error bars generated with \verb|PandExo|.

			\begin{figure*}[!ht]
				\centering
				\includegraphics[width=\hsize]{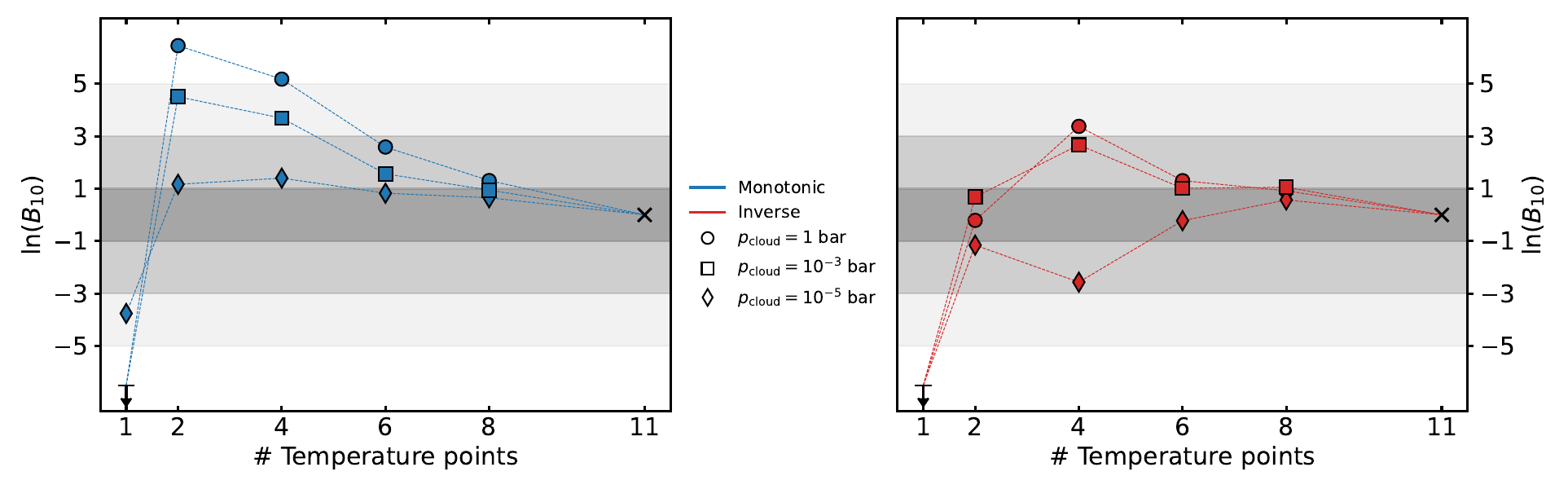}
				\caption{Same as Fig.~\protect\ref{fig:bf-drs}, but for transmission spectra generated using the transit-depth error bars resulting from \texttt{PandExo} simulations.}
				\label{fig:bf-pandexo}
			\end{figure*}
        \end{onecolumn}

        \section{Retrieval parameter accuracy}\label{sec:retrieval-accuracy-expl}
            
\begin{figure*}[!ht]
				\centering
				\includegraphics[width=\hsize]{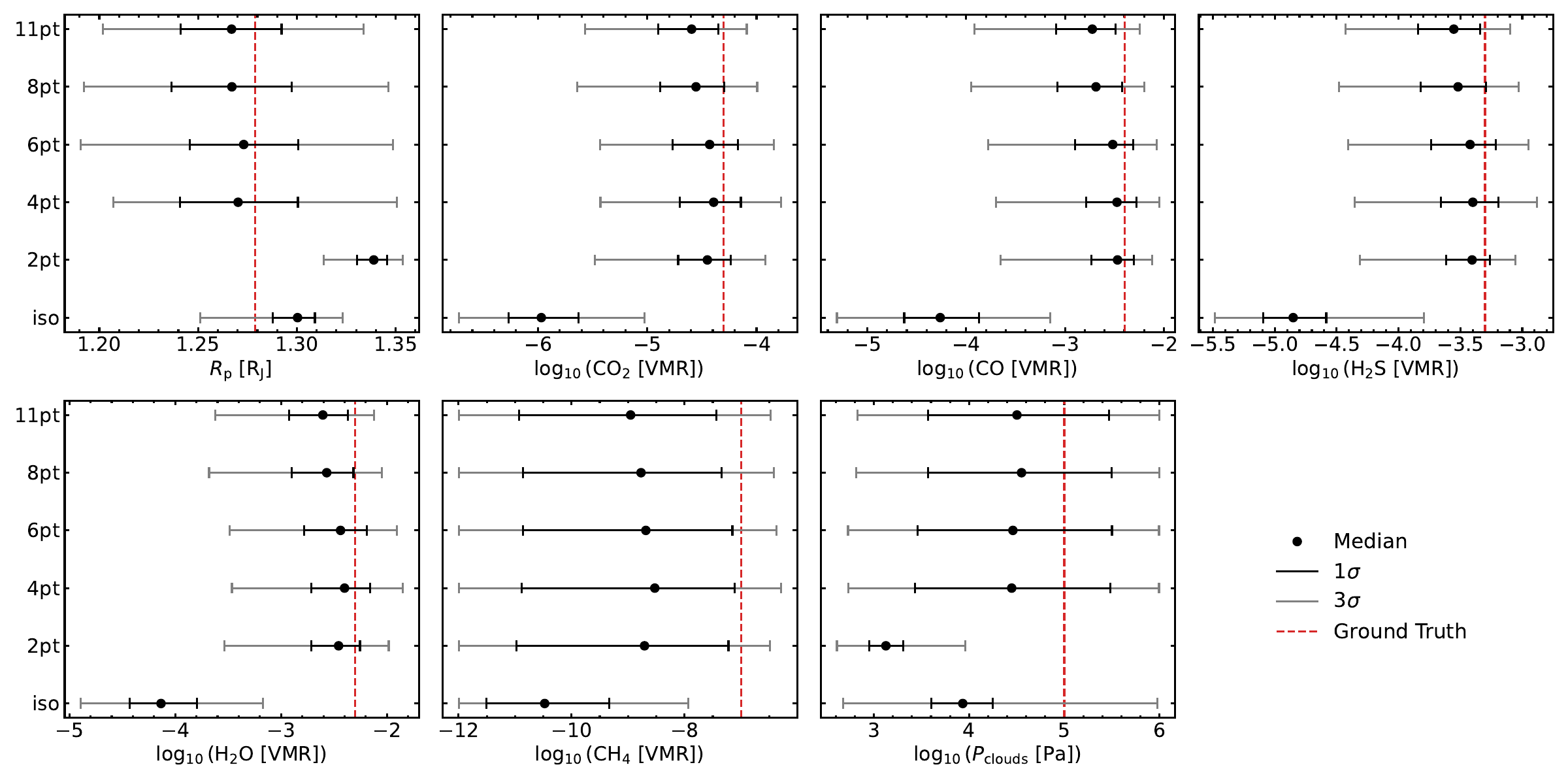}
				\caption{Accuracy of retrieved parameters, in comparison to known input value, for the synthetic transmission spectrum generated with the ``inverse'' p-T profile, a cloud-top pressure of \SI{1}{\bar}. In all panels, the parameter value is shown in the x-axis, and varying retrieval cases are listed on the y-axis. For each retrieval case, the median retrieved value (black dot), as well as the 1$\sigma$ uncertainty provided by \texttt{TauREx} (black error bar), and the boundaries of the $3\sigma$-equivalent CCI (grey error bars) are shown. The known true parameter value is indicated with a red dashed line.}
				\label{fig:parameter-accuarcy-example-drs}
			\end{figure*}

\begin{figure*}
				\centering
				\includegraphics[width=\hsize]{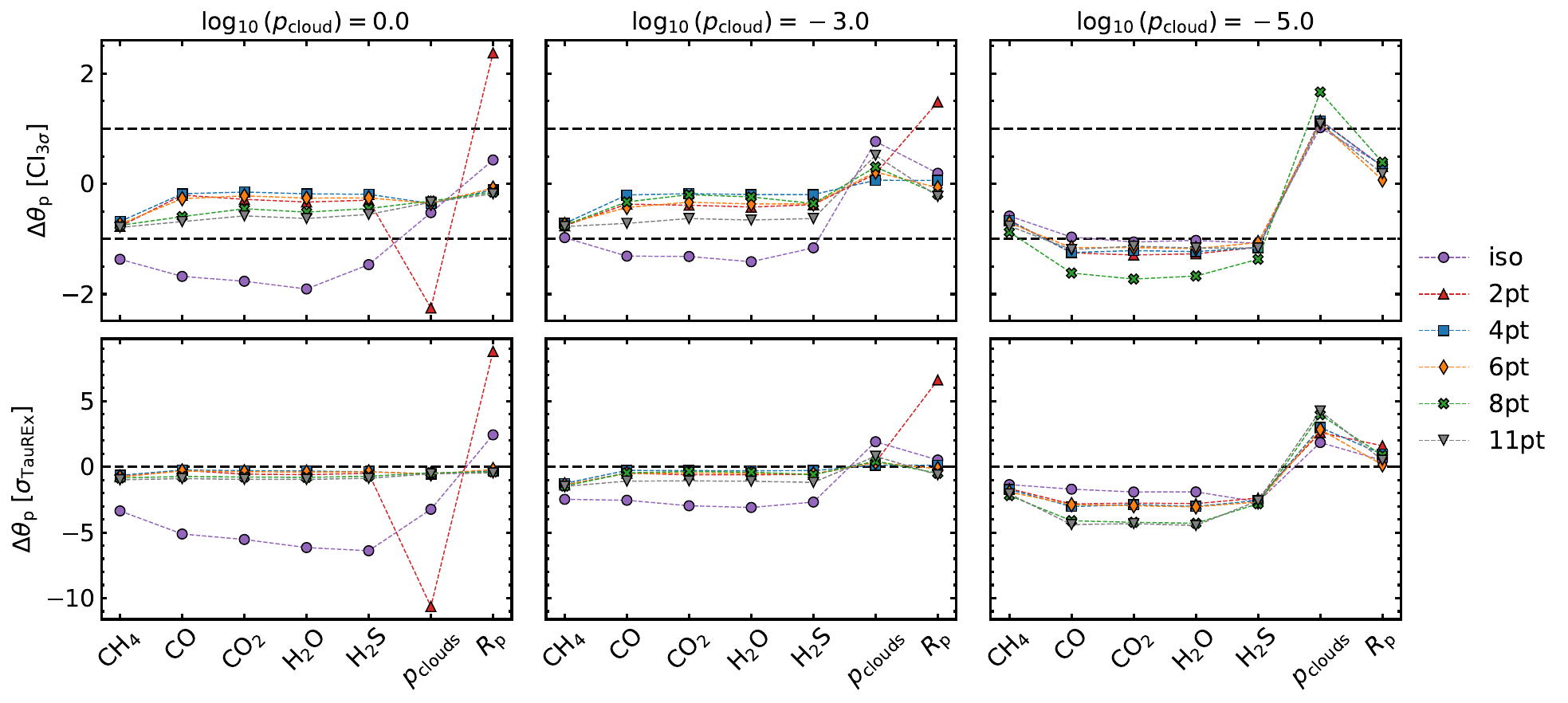}
				\caption{Accuracy of retrieved parameters for the case of the ``inverse'' input p-T profile, and DRS transit depth error bars. The individual panels show the retrieved homogeneous VMRs, cloud-top pressure ($p_\mathrm{clouds}$), and planetary radius ($R_\mathrm{p}$) on the x-axis. The top and bottom panels show the accuracy evaluation based on the 3$\sigma$-equivalent centred credible interval, and 1$\sigma$ parameter estimation uncertainty value from \texttt{TauREx}, respectively (see also Fig.~\ref{fig:parameter-accuarcy-example-drs}). The individual columns show, from left to right, cases for a cloud-top pressure of $1$ \si{\bar}, $10^{-3}$ \si{\bar}, and $10^{-5}$ \si{\bar}, respectively. Results are shown for the retrieval of an isothermal (purple), two-point (red), four-point (blue), and six-point (orange), eight-point (green), and eleven-point (grey) p-T profile, as described in Sect.~\ref{sec:results}.}
				\label{fig:parameter-accuracy-drs-inv}
			\end{figure*}

\begin{figure*}
				\centering
				\includegraphics[width=\hsize]{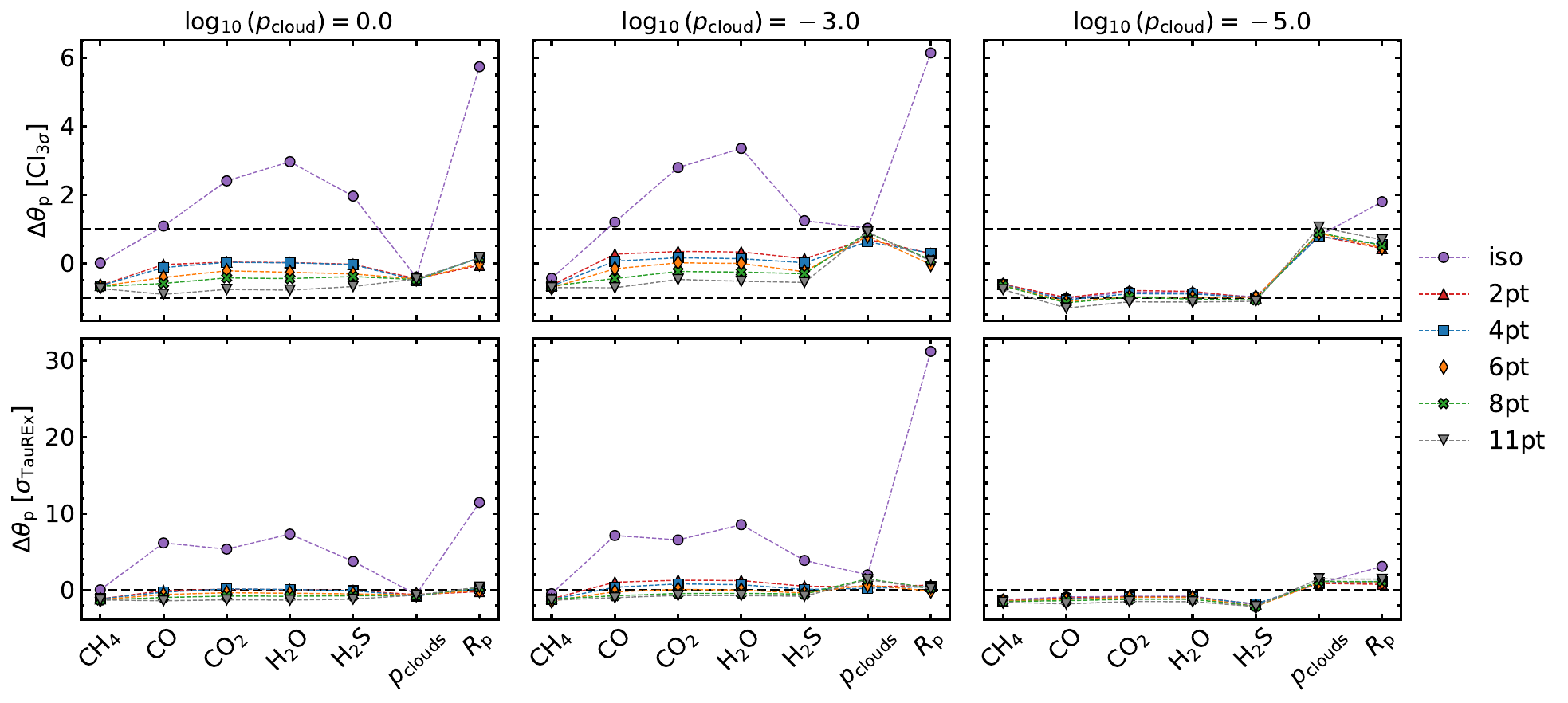}
				\caption{Same as Fig.~\ref{fig:parameter-accuracy-drs-inv}, but for the case of the ``monotonic'' input p-T profile.}
				\label{fig:parameter-accuracy-drs-norm}
			\end{figure*}

\begin{figure*}
				\centering
				\includegraphics[width=\hsize]{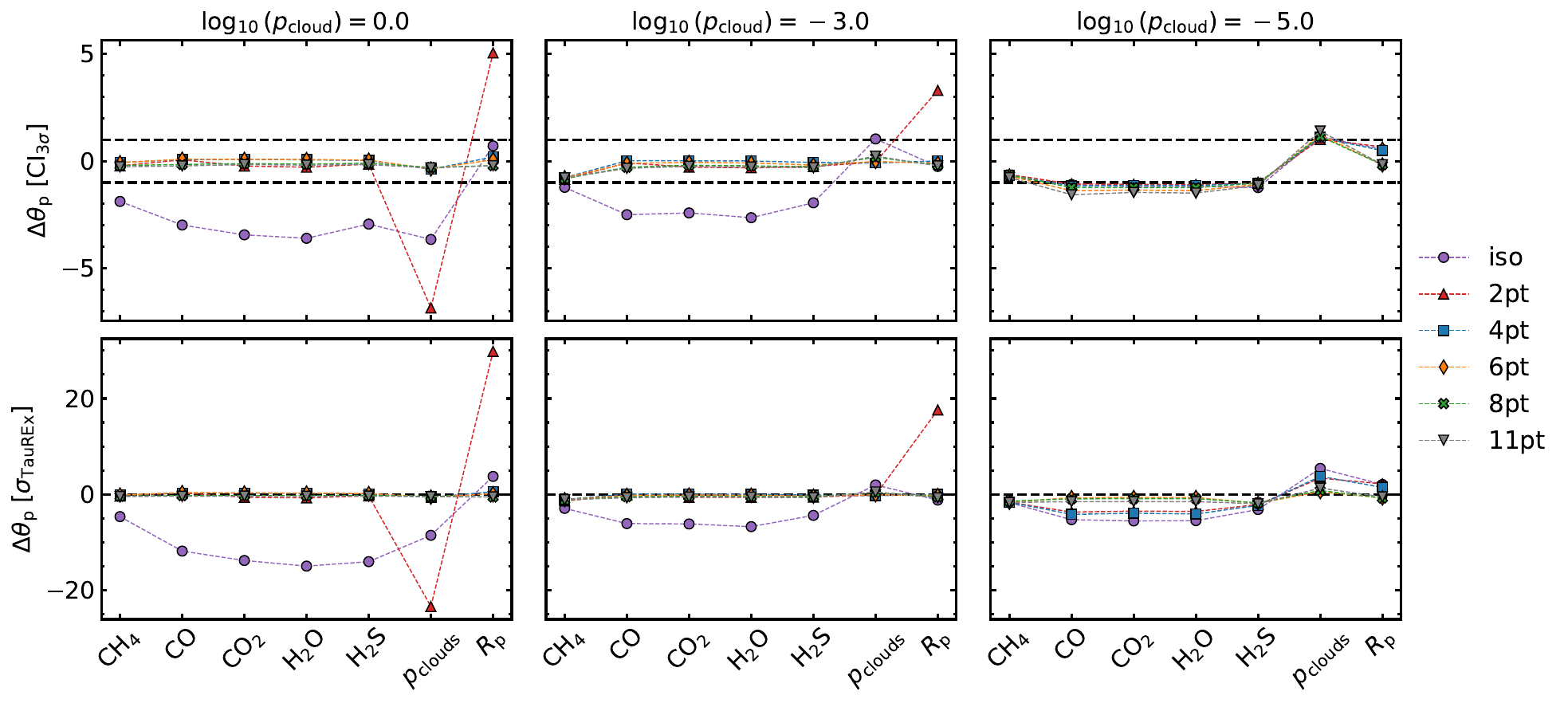}
				\caption{Accuracy of retrieved parameters for the case of the ``inverse'' input p-T profile, and \texttt{PandExo} transit depth error bars. The individual panels show the retrieved homogeneous VMRs, cloud-top pressure ($p_\mathrm{clouds}$), and planetary radius ($R_\mathrm{p}$) on the x-axis. The top and bottom panels show the accuracy evaluation based on the 3$\sigma$-equivalent centred credible interval, and 1$\sigma$ parameter estimation uncertainty value from \texttt{TauREx}, respectively. The individual columns show, from left to right, cases for a cloud-top pressure of $1$ \si{\bar}, $10^{-3}$ \si{\bar}, and $10^{-5}$ \si{\bar}, respectively. Results are shown for the retrieval of an isothermal (purple), two-point (red), four-point (blue), and six-point (orange), 8-point (green), and eleven-point (grey) p-T profile, as described in Sect.~\ref{sec:results}.}
				\label{fig:parameter-accuracy-pandexo-inv}
			\end{figure*}

\begin{figure*}
				\centering
				\includegraphics[width=\hsize]{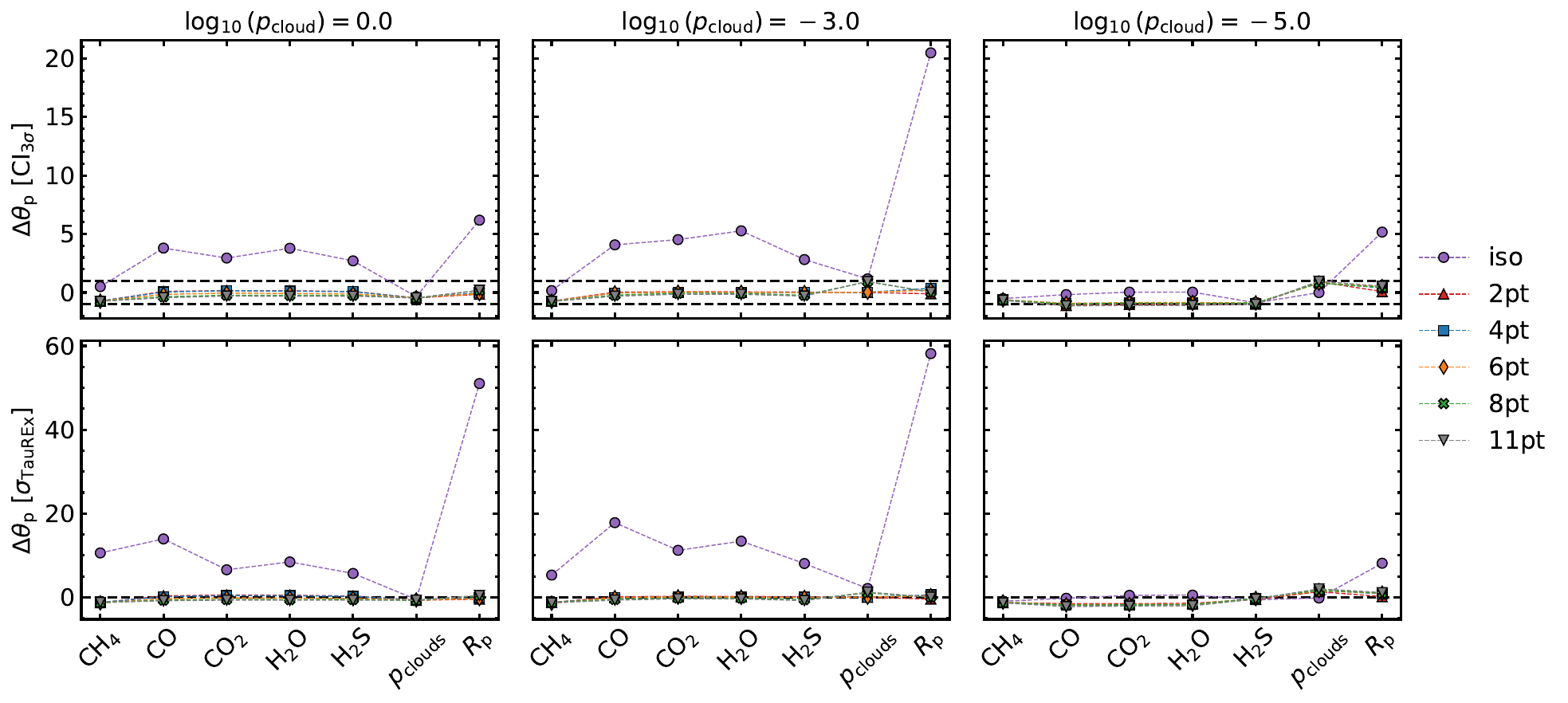}
				\caption{Same as Fig.~\ref{fig:parameter-accuracy-pandexo-inv}, but for the case of the ``monotonic'' input p-T profile.}
				\label{fig:parameter-accuracy-pandexo-mon}
			\end{figure*}

            Supplementary to the parameter accuracy evaluation discussed in Sect.~\ref{sssec:parameter-accuracy}, we show an example of the 3-sigma-equivalent centred credible interval and the 1$\sigma$ parameter uncertainty reported by \texttt{TauREx} in Fig.~\ref{fig:parameter-accuarcy-example-drs}. Figures~\ref{fig:parameter-accuracy-drs-inv} - \ref{fig:parameter-accuracy-pandexo-mon} summarise the accuracy evaluation for all synthetic transmission spectra considered in this work\footnote{We refer to the supplementary repository for a full set of all individual accuracy evaluations: \url{https://doi.org/10.5281/zenodo.13737625}}.
            
            \FloatBarrier
        
        \begin{onecolumn}
            \section{Additional retrieved p-T profiles}\label{sec:additional-pt-profile}
            
            Complementary to Fig.~\ref{fig:accuracy-pt-profiles-drs}, we show the retrieved isothermal, two-point, and four-point p-T profiles for all input transmission spectra generated with the ``Pandexo'' noise case in Fig.~\ref{fig:accuracy-pt-profiles-pandexo}, together with the underlying input pressure-temperature profiles, and the pressure-layer of the associated flat-opacity cloud deck.
            
            \begin{figure*}[!ht]
            
				\centering
				\includegraphics[width=\hsize]{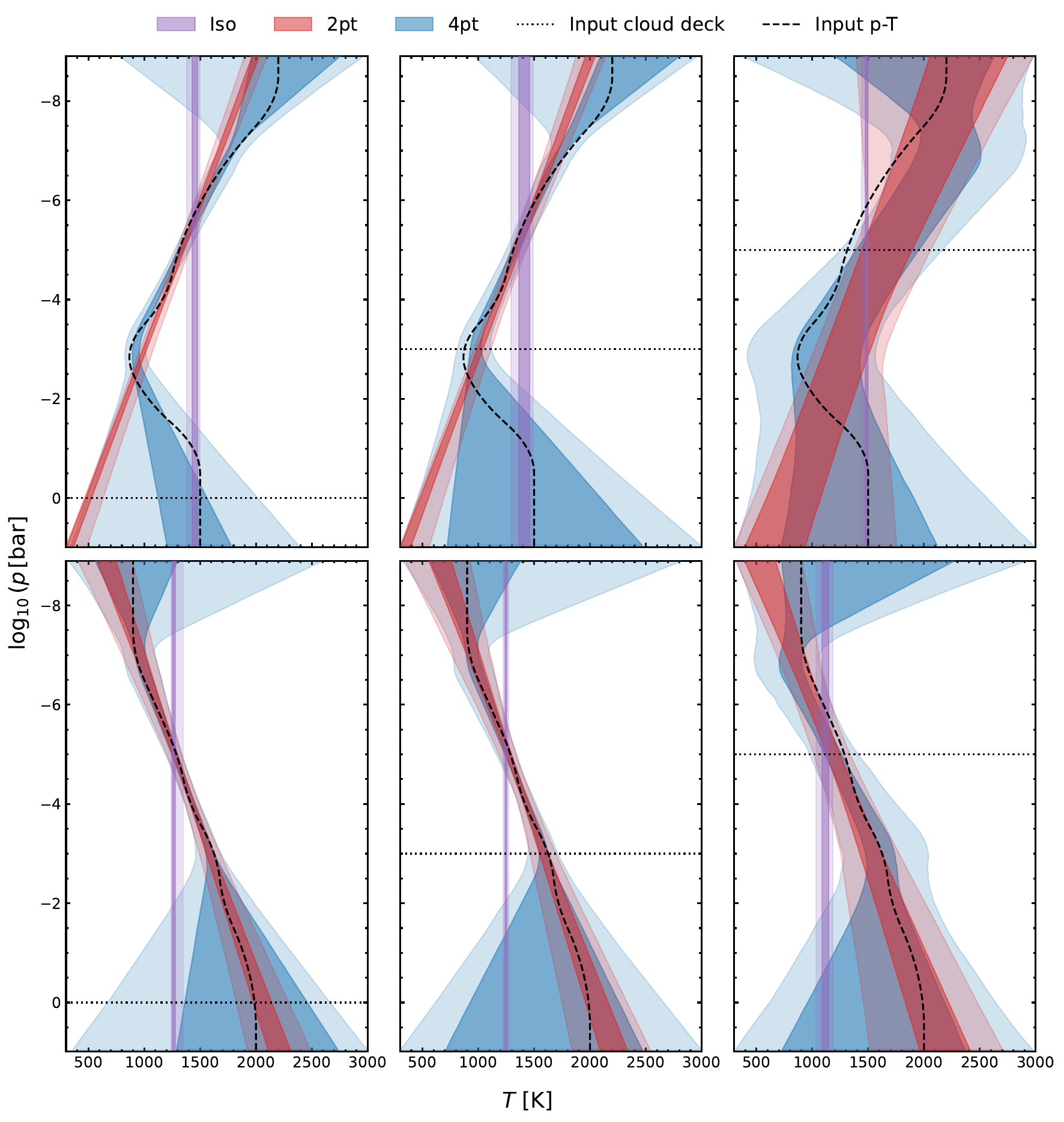}
				\caption{
					Same as Fig.~\ref{fig:accuracy-pt-profiles-drs}, but for the synthetic spectra generated using the ``Pandexo'' noise case.
				}
				\label{fig:accuracy-pt-profiles-pandexo}
			\end{figure*}
		\end{onecolumn}

	\end{appendix}

\end{document}